\documentclass{aa} 

\usepackage{graphicx}
\usepackage{txfonts}
\usepackage[version=4]{mhchem}
\usepackage{amsmath}
\usepackage{subcaption}
\usepackage{booktabs}
\usepackage{xcolor}

\usepackage[switch]{lineno} 
\usepackage[hidelinks]{hyperref}

\begin{document}

   \title{Interstellar Formation of Thioethanal (CH$_3$CHS)
   \thanks{Revised manuscript submitted to Astronomy \& Astrophysics.}}
   \subtitle{Gas-Phase and Ice-Surface Mechanisms involving Secondary Sulfur Products}

\author{N. Rani\inst{1} 
        \and S. Vogt-Geisse\inst{1}
        \and S. Bovino\inst{2,3,4}
        \thanks{Corresponding authors: NR (\email{nrani@udec.cl}) and SVG (\email{stvogtgeisse@qcmmlab.com}).}
       }

\institute{
Departamento de Físico-Química, Facultad de Ciencias Químicas, Universidad de Concepción, Concepción, Chile 
\and
Department of Chemistry, University of Rome Sapienza, P.le A. Moro 5, 00185, Rome, Italy
\and
Departamento de Astronom\'ia, Facultad Ciencias F\'isicas y Matem\'aticas, Universidad de Concepci\'on, Av. Esteban Iturra s/n Barrio Universitario, Casilla 160, Concepci\'on, Chile
\and
INAF, Osservatorio Astrofisico di Arcetri, Largo E. Fermi 5, I-50125, Firenze, Italy
}

 
  \abstract
   {The formation pathways of sulfur-bearing species in the interstellar medium are crucial for understanding astrochemical processes in cold molecular clouds and to gain new insights about the sulfur budget in these regions.}
   {We aim to explore the recently detected, thioethanal (CH$_3$CHS) formation mechanisms from thioethanol (CH$_3$CH$_2$SH) as a precursor in addition to secondary sulfur products.}
   {We employ electronic structure methods and density functional theory for both gas-phase and ice-grain surface environments. To mimic interstellar ice-mantles, we use both medium (W6) and large amorphized (W22) water clusters as implemented in Binding Energy Evaluation protocol.}
   {We identify a barrierless formation mechanism for CH$_3$CHS that remains kinetically feasible under low-temperature interstellar conditions, in the gas-phase. \textcolor{black}{Surface environments modulate activation barriers in a site-specific manner, elucidated through both Langmuir-Hinshelwood and Eley-Rideal initiated surface reaction pathways.} Compared to oxygen analogs, sulfur chemistry enables alternate pathways due to weaker S–H bonding, with a competing route forming ethane-1,1-di-thiol (CH$_3$CH(SH)SH), \textcolor{black}{on the ice-grain surface,} potentially reducing CH$_3$CHS yields.The first accurate binding energy for thioethanol on water ice is also reported, confirming its greater volatility than ethanol.} 
   { The proposed mechanism offers a \textcolor{black}{tentative hypothesis for the apparent mutual exclusive detections of the CH$_3$CH$_2$SH and CH$_3$CHS in TMC-1, Orion, and Sgr B2(N), that further requires validation through quantitative astrochemical modeling and also to distinguish this chemical differentiation from observational sensitivity limitations.} These \textcolor{black}{qualitative} findings highlight the multifaceted chemical behavior of sulfur-bearing organics in the interstellar medium and support CH$_3$CH(SH)SH as promising astrochemical targets.}

    \keywords{ISM: molecules -- Molecular Data -- Astrochemistry -- methods: numerical}

   \maketitle
   \nolinenumbers
%

\section{Introduction} \label{sec:intro}

The interstellar medium (ISM) is a chemically rich and dynamic environment where sulfur plays a vital role, yet its astrochemical behavior remains elusive due to the long-standing sulfur depletion problem \citep{tielens2005physics}. Although sulfur’s cosmic abundance is estimated at $\sim$10$^{-5}$ relative to hydrogen, observations show it is depleted by over two orders of magnitude in cold, dense molecular clouds compared to diffuse regions \citep{Anderson_2013}. The detection of the first S-bearing molecule, CS, in the ISM by \citet{penzias1971interstellar} initiated decades of exploration into sulfur chemistry. However, only SO$_2$ and OCS have been securely identified in interstellar ices, leaving the bulk sulfur reservoir unknown \citep{goicoechea2006low, fuente2019gas}. Updated astrochemical models have rejected the idea of sulfur being primarily atomic in these environments \citep{vidal2017reservoir, laas2019modeling}, instead suggesting that sulfur may be sequestered in complex organic molecules within dust grains. This view is supported by cometary studies of 67P/Churyumov–Gerasimenko, where abundant organosulfur species are detected without evidence of depletion, implying preservation of interstellar sulfur in solid form \citep{rubin2019volatile}. Resolving the molecular identity of this hidden sulfur is essential to understanding chemical evolution in star- and planet-forming regions.\\
Moreover, sulfur and oxygen share chemical similarities, with sulfur exhibiting a broader range of oxidation states, which leads to the formation of various molecular species. Despite this, the observed abundance of sulfur-containing molecules in dense regions of cold molecular clouds is lower than predicted based on the sulfur-to-oxygen ratio \citep{McGuire_2022}. This discrepancy suggests that many sulfur-bearing molecules remain undiscovered in the ISM, as indicated by comparisons of interstellar censuses of oxygen and sulfur compounds. Complex sulfur molecules, such as thiol (CH$_3$CH$_2$SH), are not unprecedented in interstellar chemistry \citep[see for example][for observations in the Taurus Molecular Cloud]{cernicharo2021tmc}.
In view of this and in the quest to understand the missing sulfur in the ISM, this study investigates the formation of thioethanal (CH$_3$CHS).
The ubiquitous presence of its oxygen analogue, acetaldehyde (CH$_3$CHO), in various astrophysical environments, ranging from cold prestellar cores to hot corinos and young discs \citep{oberg2010cold, bacmann2012detection, codella2015astrochemistry} has long made CH$_3$CHS a promising target for detection.\\
 While previous observational efforts failed to identify CH$_3$CHS in sources such as the hot molecular core SgrB2(N2), as well as in various prestellar and protostellar sources near low-mass star-forming regions ~\cite{margules2020submillimeter}, its recent detection in the cold dark cloud TMC-1 marks a significant milestone in interstellar sulfur chemistry ~\citep{agundez2025detection}.
 In addition to this detection, astrochemical modeling is performed to explore the possible formation of CH$_3$CHS, incorporating the reaction between ethyl radical (C$_2$H$_5$) and atomic sulfur (S) as a key formation pathway ~\citep{agundez2025detection}. However, the predicted abundance from this model is lower than the observed abundance of CH$_3$CHS, suggesting that additional formation routes must be contributing to its presence. 
 Recent experimental work by ~\citet{santos2024formation} tentatively detected CH$_3$CHS in interstellar ice experiments at 10 K, proposing a proton relay-type isomerization as a possible formation mechanism, despite its high energy barriers. Additionally, ~\citet{purzycka2021uv} identified CH$_3$CHS as a major organic sulfur molecule resulting from the UV photolysis of thioethanol. However, a comprehensive theoretical understanding of CH$_3$CHS formation remains lacking, necessitating further investigation into its potential reaction pathways.\\
The primary aim of this study is to determine whether thioethanol (CH$_3$CH$_2$SH) can serve as a precursor for the thermal formation of thioethanal (CH$_3$CHS) within cold molecular clouds. The objectives of this study are two fold: firstly, to examine the formation 
pathways of CH$_3$CHS, and secondly, to determine whether the results align with those of its oxygen analogue, thereby assessing the reliability of employing the kinetic data in astronomical modeling through comparative 
analysis. \textcolor{black}{In addition to the primary pathways, equation \eqref{eq:reaction1}, \eqref{eq:reaction2a}, \eqref{eq:reaction2b}, leading to (CH$_3$CHS), we also consider the formation of ethane-1,1-di-thiol (CH$_3$CH(SH)SH) as a competing branch (equation 2c), given the astrochemical relevance of atomic hydrogen.} The reaction mechanism under consideration is:
\begin{equation}
    \ce{CH3CH2SH + OH -> CH3CHSH + H2O}
    \label{eq:reaction1}
\end{equation}
\begin{subequations}
\begin{equation}
    \ce{CH3CHSH + S -> CH3CHSSH }
    \label{eq:reaction2a}
\end{equation}
\begin{equation}
    \ce{CH3CHSSH -> CH3CHS + SH}
    \label{eq:reaction2b}
\end{equation}
\begin{equation}
    \textcolor{black}{\ce{CH3CHSSH + H -> CH3CH(SH)SH}}
    \label{eq:reaction2c}
\end{equation}
\end{subequations}

Thioethanol is detected in Orion KL, Sgr B2(N2) \citep{muller2016exploring, rodriguez2021thiols, kolesnikova2014spectroscopic} in the gas phase in \textit{cis} conformation, with tentative detections of anti-thioethanol. Interestingly, CH$_3$CH$_2$SH and CH$_3$CHS exhibit complementary detection patterns across sources like TMC-1, Orion, and Sgr B2(N) \citep{agundez2025detection}. Further, the OH radical is a common interstellar species, believed to form in ice mantles via photodissociation, radiolysis, or atom addition. Atomic sulfur, although its role in the sulfur budget is debated, may exist in low concentrations due to photodissociation and radiolysis in regions with lower visual extinction values. This is supported by \citet{purzycka2021uv}, confirming the presence of sulfur atoms in their experimental findings considering CH$_3$CH$_2$SH as the primary reactant in the ices.\\
Given the presence of these reactants in cold molecular clouds, we employed state-of-the-art quantum chemical methods to study the reaction barriers in both the gas phase and on ice dust surfaces. Interstellar ices primarily exist in an amorphous state, rather than as crystalline structures, forming a heterogeneous network of hydrogen bonds (H-bonds) that significantly influences molecular interactions \citep{hama2013surface, he2011interaction, noble2012thermal}. The strength of these interactions, quantified by the binding energy (BE), determines molecular residency times on the surface and consequently affects their availability for reactive encounters. Experimental and theoretical studies have demonstrated that the distribution of BE values for a given species across a representative set of binding sites follows a Gaussian-like profile \citep{grassi2020novel, bovolenta2020high}. Furthermore, the intricate H-bonding environment can modulate surface reactions by either facilitating or inhibiting specific pathways. To model this complexity and accurately describe reactive binding sites, we utilized the Binding Energy Evaluation Platform (BEEP), an in-house computational framework designed to capture the diverse adsorption environments present in amorphous solid water (ASW) \citep{bovolenta2022binding}. The binding energy distributions of reactant is computed to study the interactions with the interstellar ice-dust to further assess its reactivity on the \textcolor{black}{ASW} surface. \\
The following sections describe the computational methodology employed, followed by the results, discussion, and astrophysical implications.
\begin{figure*}[tbp]
    \centering
    \includegraphics[width=0.95\linewidth]{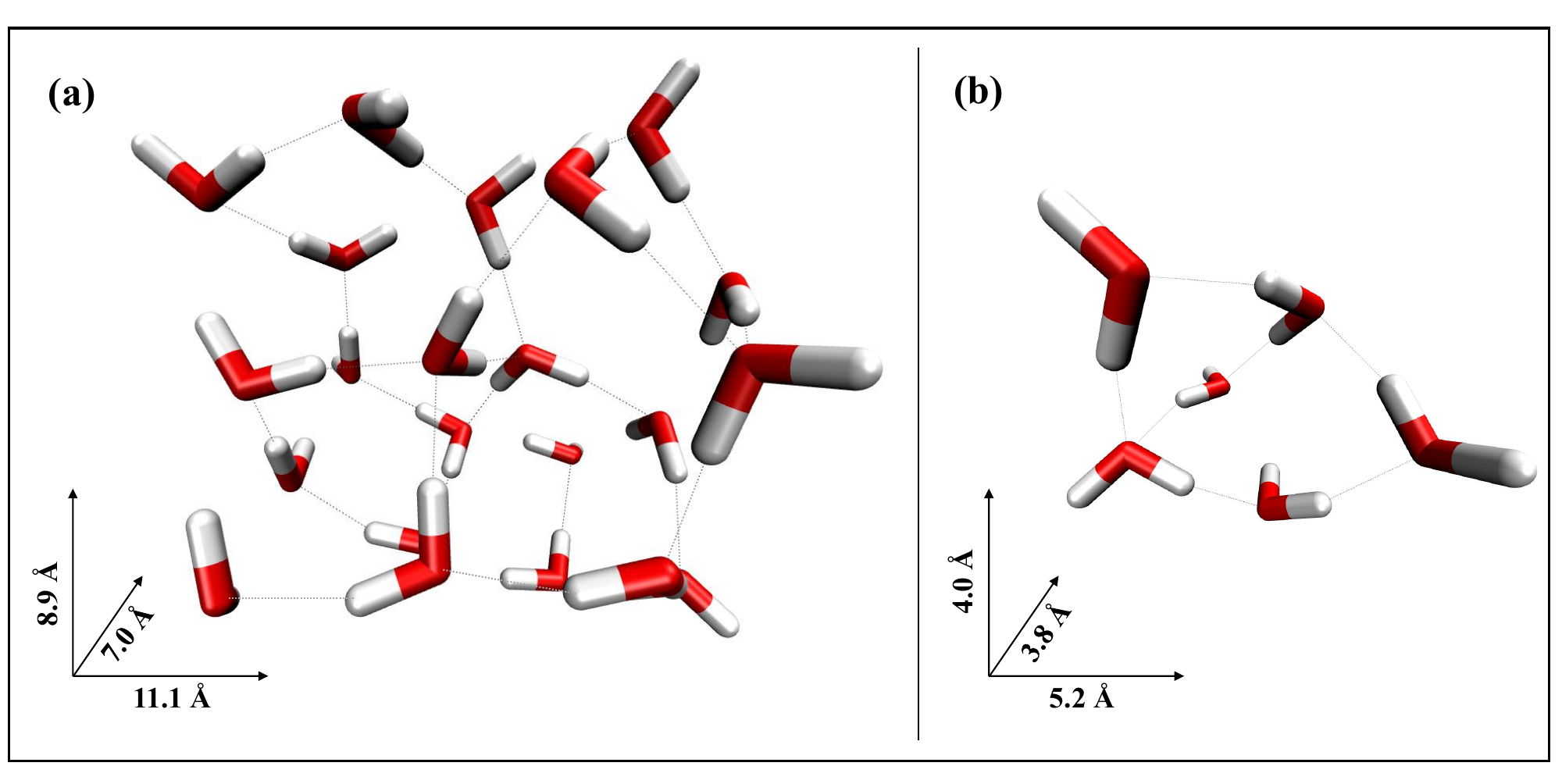}\\
    \caption{\textcolor{black}{Representative structures of the water clusters used to investigate binding energies and reactivity. (a) One of the seventeen W22 clusters; (b) one of the three W6 clusters. Approximate dimensions are in \AA{}. Oxygen is in red and hydrogen in white.}}.
    \label{fig:water_cluster}

\end{figure*}
\section{Computational Methods} \label{sec:method}
\subsection{Gas-phase computations} \label{Gas-phase mechanism methodology}
To investigate the gas-phase reaction mechanism, we initially employed the cost-effective BHandHLYP/def2-TZVP approach, followed by geometry refinement at the \textcolor{black}{M06-2X/def2-TZVP } level of theory ~\citep{becke1993new,zhao2008m06,weigend2005balanced,woon1993gaussian}. The choice of functionals is guided by an extensive benchmarking study against high-level DF-CCSD(T)-F12/cc-pVDZ-F12 reference calculations, ensuring reliable structural accuracy ~\citep{peterson2008systematically, knizia2009simplified, deprince2013accuracy}. The detailed description of the benchmarking procedure and its results is provided in Fig. \ref{fig:benchmarking}, Appendix A. 
Further, all stationary points are characterized as either first-order saddle points or local minima through vibrational frequency analysis at the corresponding levels of theory. Transition states are further validated using Intrinsic Reaction Coordinate (IRC) calculations, as implemented in the Gaussian16 program suite ~\citep{g16}.
The barrierless nature of the reaction steps involving open-shell species is examined using  broken-symmetry Density Functional Theory (DFT) approach~\citep{neese2004definition},
\textcolor{black}{by placing the two radical fragments at a separation of 4~\AA{} to avoid artificial bonding. Additionally relaxed potential energy scans were carried out along the new bond formation on both at the low-spin and high-spin electronic states employing M06-2X/def2-TZVP, see Appendix \ref{appendix:ScanSection} for more details.}\\
Final energies are computed at the CCSD(T)/CBS level of theory utilizing the PSI4 program~\citep{turney2012psi4}. These include zero-point vibrational energy (ZPVE) correction at \textcolor{black}{M06-2X/def2-TZVP} level of theory.\\

\subsection{Ice-dust grain mechanism methodology} \label{subsec:Ice-grain mechanism methodology}
For studying reaction channels on ice-dust grains, we employed a model consisting of clusters with 22 water molecules, which is integrated into BEEP \footnote{\url{https://github.com/QCMM/BEEP}}, powered by QCArchive software suite~\citep{smith2021molssi}. This model features 17 distinct amorphous water clusters, derived from \textit{ab initio} molecular dynamics (MD) simulations of a single 22-water-molecule cluster (W22) \citep{bovolenta2020high}, \textcolor{black}{one of the representative structure is shown in Fig. \ref{fig:water_cluster}(a)}. Further details of the simulation methodology are available in the work by \citet{bovolenta2022binding}. Each cluster captures different segments of the ASW surface, offering a diverse range of binding sites. Consequently, this comprehensive set of clusters presents numerous potential locations for reactive interactions.\\
\textcolor{black}{In addition, BEEP includes three medium-sized ASW clusters containing six water molecules (W6), Fig.~\ref{fig:water_cluster}(b), also used in this work, detailed in section \ref{subsec:Binding Sites} and \ref{Selection of Reactive Binding Sites}}.\\
To explore the different types of binding modes, the primary reactant, CH$_3$CH$_2$SH, is sampled on the W22 
set-of-clusters, resulting in the identification of 130 distinct binding sites using the HF-3c/MINIX level of theory ~\citep{sure2013corrected}. \textcolor{black}{The reliability of HF-3c/MINIX method is discussed in Appendix ~\ref{appendix:DFT} and depicted in Fig. \ref{fig:benchmarkinghf3c}. The MPWB1K-D3(BJ)/def2-TZVPD functional ~\citep{zhao2004hybrid, weigend2005balanced, grimme2011effect} is selected for evaluating the binding energies of all sampled sites, ~\textcolor{black}{based on the benchmarking with CCSD(T)/CBS reference binding energy, represented in Appendix ~\ref{appendix:DFT}, Fig.~\ref{fig:benchmarkingW1}} }. The binding site optimizations in BEEP are performed using TERACHEM and PSI4 v1.6 software, with PSI4 also utilized for calculating binding energies ~\citep{seritan2021terachem, smith2021molssi}. The binding energy (\(\Delta E_{\text{bind}}\)) of CH$_3$CH$_2$SH adsorbed on an ice surface is calculated as: 

\begin{equation}
\Delta E_{\text{bind}} = E_{\text{CH}_{3}\text{CH}_{2}\text{SH}\_{\text{W22}}} - \left(E_{\text{W22}} + E_{\text{CH}_{3}\text{CH}_{2}\text{SH}}\right)
\end{equation}

\noindent where \(E_{\text{CH}_{3}\text{CH}_{2}\text{SH}\_{\text{W22}}}\) represents the total energy of the adsorbate-ice complex, \(E_{\text{W22}}\) corresponds to the energy of the isolated ice cluster, and \(E_{\text{CH}_{3}\text{CH}_{2}\text{SH}}\) is the energy of the free adsorbate in the gas phase.
By convention, \(\Delta E_{\text{bind}}\) is reported as a positive quantity, indicating the strength of the adsorption interaction. To improve the accuracy of these calculations, all reported binding energies are corrected for ZPVE contributions and include counterpoise (CP) corrections to mitigate BSSE ~\citep{bovolenta2022binding}. Sampling binding sites on the set-of-clusters, results in a distribution of \(\Delta E_{\text{bind}}\) values, which provides a more realistic representation of desorption behavior from ASW surfaces and accounts for the heterogeneous nature of interstellar ice environments. Detailed information regarding the binding energy computations is available in ~\citet{bovolenta2020high, bovolenta2022binding}.\\
\begin{figure*}[tbp]
    \centering
    \includegraphics[width=\linewidth]{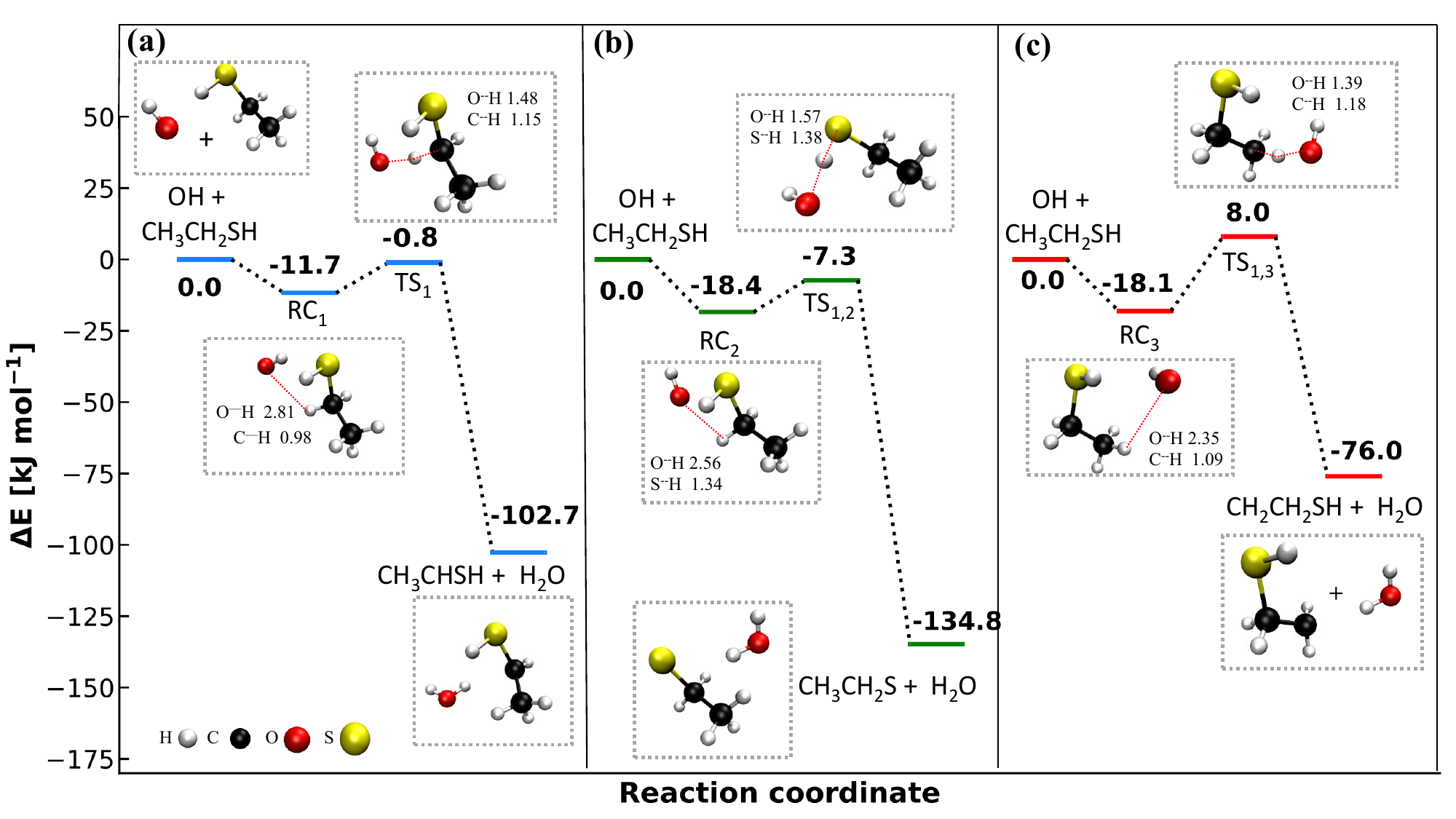}\\
    \caption{Three possible pathways for hydrogen abstraction from CH$_3$CH$_2$SH by the OH radical: (a) abstraction from the -CH$_2$ group, (b) abstraction from the -SH group, and (c) abstraction from the -CH$_3$ group. All geometries are optimized using the \textcolor{black}{M06-2X/def2-TZVP} level of theory, with energies calculated at CCSD(T)/CBS, including zero-point energy (ZPVE) corrections at \textcolor{black}{M06-2X/def2-TZVP}. The ... represent bond formation and cleavage processes, with significant bond lengths indicated in \AA{}. The atom color scheme is consistent throughout the paper and is illustrated separately for reference.}
    \label{fig:step1}

\end{figure*}
Further, for reaction mechanism two interaction scenarios are considered ~\textcolor{black}{for the initiation step}~\ref{eq:reaction1}: (i) OH interacting directly with the ASW cluster alongside the primary reactant, and (ii) OH interacting solely with the primary reactant from the gas phase. The former scenario is used to explore the ~\textcolor{black}{Langmuir-Hinshelwood (LH) initiated pathways, while the latter is investigated to understand the Eley-Rideal (ER) initiated pathways ~\citep{herbst2017synthesis}}. \textcolor{black}{In the subsequent steps, \eqref{eq:reaction2a} and in \eqref{eq:reaction2c} the atomic S and H respectively are considered to react from the gas phase. Notably, for the first reaction step in LH-initiated pathways, the energies are reported relative to the CH$_3$CH$_2$SH and OH both co-adsorbed on the cluster and for ER-initiated pathways, relative to the CH$_3$CH$_2$SH adsorbed on the cluster and OH in the gas phase at infinite separation. For subsequent steps, the energy reference corresponds to the adsorbed intermediate and the second reactant in the gas-phase.}\\
Transition states are identified using the Berny optimization algorithm implemented in Gaussian 16, ~\citep{g16} followed by frequency analysis and IRC calculations at the BHandHLYP/def2-TZVP level of theory and \textcolor{black}{the resulting geometries were subsequently refined at the M06-2X/def2-TZVP level of theory. Single-point energies were then evaluated at the same level used for binding-energy calculations, namely MPWB1K-D3(BJ)/def2-TZVPD, which has been shown to perform well for activation energies relative to CCSD(T)/CBS//M06-2X/def2-TZVP benchmark values for representative gas-phase reactions (see Appendix Fig. \ref{fig:bnchmark_rxnEnergy}).}

\begin{figure*}[tbp]
    \centering
    \includegraphics[width=\linewidth]{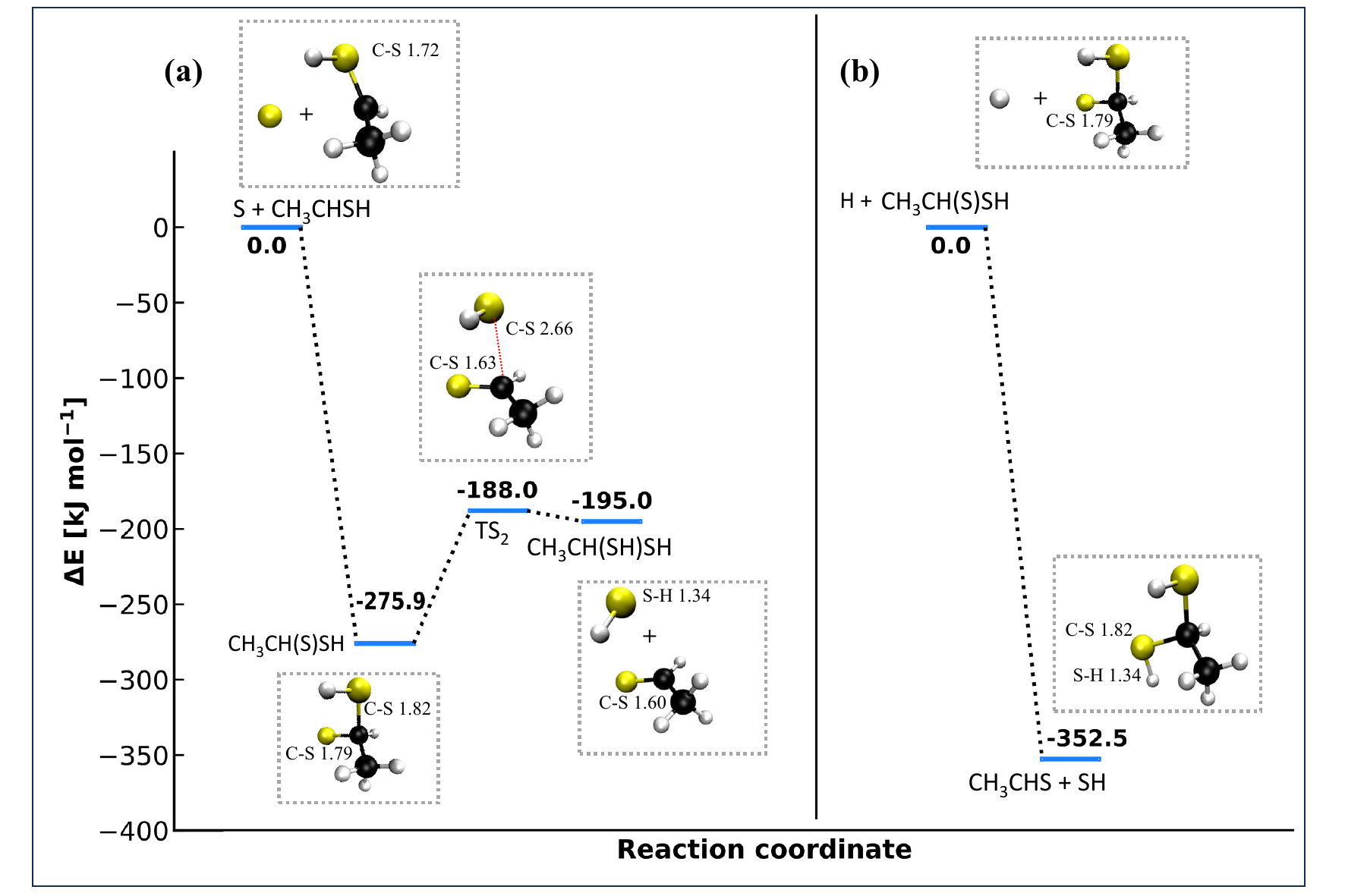}\\
    \caption{\textcolor{black}{Gas-phase reaction mechanism: (a) Step 2a, 2b and (b) Step 2c, CH$_3$CH(S)SH + H recombination, is shown only for comparison with the surface mechanism and is not expected to be feasible in the gas phase. Geometries are optimized at the M06-2X/def2-TZVP  level of theory, and energies are computed using CCSD(T)/CBS with zero-point energy (ZPVE) corrections at \textcolor{black}{M06-2X/def2-TZVP }. Significant bond lengths are shown in \AA{}}}
    \label{fig:gas-phase}

\end{figure*}

\section{Results} \label{sec:Results}
\subsection{Gas-phase Reaction Mechanism} \label{subsec:Gas-phase Reaction Mechanism}
The gas-phase mechanism for the conversion of thioethanol (CH$_3$CH$_2$SH) to thioethanal (CH$_3$CHS) involves multiple steps initiated by the attack of OH radicals (see equation \eqref{eq:reaction1}). There are three possible hydrogen abstraction pathways, methylene (-CH$_2$) hydrogen, thiol group hydrogen (-SH) and methyl (-CH$_3$) hydrogen. The respective energy profile at \textcolor{black}{CCSD(T)/CBS//M06-2X/def2-TZVP } level of theory 
is shown in Fig.~\ref{fig:step1}. \textcolor{black}{The OH radical abstracts a methylene hydrogen from thioethanol, Fig.~\ref{fig:step1}(a), yielding radical CH$_3$CHSH and H$_2$O. This reaction proceeds through a weakly bound reaction complex (RC$_1$) with an energy of –11.7 kJ/mol relative to the reactants (CH$_3$CH$_2$SH + OH) at \textcolor{black}{{CCSD(T)/CBS//M06-2X/def2-TZVP }} level of theory. The transition state (TS$_1$) lies at \textcolor{black}{–0.8} kJ/mol, yielding a submerged barrier with a height of \textcolor{black}{10.9} kJ/mol. The product formation is exothermic, with the CH$_3$CHSH + H$_2$O lying at –102.7 kJ/mol. The alternative -S(H) abstraction is also feasible at low temperatures as it presents
a submerged TS that lies  7.3 kJ/mol below the isolated reactants., Fig.~\ref{fig:step1}(b). However, the synthesis of thioethanal (CH$_3$CHS) from the radical CH$_3$CH$_2$S is reported to be associated with a high activation energy barrier ~\citep{agundez2025detection}. Furthermore, the methyl -H is the least favorable both kinetically as well as thermodynamically, Fig.~\ref{fig:step1}(c). Consequently, we focus on the pathway of the extraction of methylene hydrogen, which allows the formation of the key intermediate, CH$_3$CHSH, and ultimately thioethanal (CH$_3$CHS). Nevertheless, these competing pathways will influence the branching ratios and will be included in further detail in our future astrochemical modeling efforts.}\\
\textcolor{black}{Further, Fig. \ref{fig:gas-phase} shows the subsequent steps of the equation \eqref{eq:reaction2a}-\eqref{eq:reaction2c}), for the formation of CH$_3$CHS in the gas-phase.} \textcolor{black}{The radical intermediate CH$_3$CHSH undergoes strongly exothermic addition with atomic sulfur (S) in its ground-state triplet form and proceeding with no discernible barrier along the C–S association coordinate (Appendix Fig.\ref{fig:scan}(a)) forming a 'thio-hemiacetal' like radical intermediate, CH$_3$CH(S)SH. The excess energy of the thio-hemiacetal radical intermediate, \textcolor{black}{–275.9 kJ/mol} relative to the reactants (CH$_3$CHSH + S), can dissociate it further,} yielding thioethanal (CH$_3$CHS) and an SH radical. The transition state (TS$_2$) for this dissociation lies at \textcolor{black}{–188.8 kJ/mol} relative to the initial reactants (CH$_3$CHS + S), with a barrier height of \textcolor{black}{87.1 kJ/mol}. Despite this high barrier, the reaction remains submerged and highly exothermic, with the final products (CH$_3$CHS + SH) lying at \textcolor{black}{–195.0 kJ/mol}. Notably, the dissociation back to reactants is chemically less favourable due to the presence of thiocarbonyl bond, -C=S. \\
In principle, the energy released during sulfur addition \textcolor{black}{(–275.9 kJ/mol)} could be sufficient to overcome the barrier \textcolor{black}{(87.1 kJ / mol)}, particularly under non-statistical or prompt dissociation conditions, where the molecule dissociates before the internal energy is fully redistributed. However, in the dilute and low-temperature environment of cold dense clouds such as TMC-1, where third-body collisions are rare and radiative energy loss is inefficient, the internal energy is likely to become randomized if not released via dissociation modes. This may increase the likelihood that CH$_3$CH(S)SH becomes kinetically stabilized and long-lived. If the radical CH$_3$CH(S)SH is kinetically trapped, an alternative pathway involves the radical-radical recombination between CH$_3$CH(S)SH and H radical, forming ethane-1,1-di-thiol (CH$_3$CH(SH)SH. This pathway is even more exothermic than the dissociation channel, leading to the formation of the most stable product across the extended reaction network involving hydrogen addition (see Fig.~\ref{fig:gas-phase}). The recombination occurs spontaneously, with no transition state ~\textcolor{black}{along the reaction coordinate}, driven by the high reactivity of the radicals, ~\textcolor{black}{Appendix \ref{fig:scan}(b)}. But is unlikely to happen in the gas phase since it requires a source of excess energy dissipation or will dissociate back to the reactants.Nevertheless, this H-recombination channel is included in Fig. \ref{fig:gas-phase} solely for comparison with the surface mechanism. Of note, in the gas-phase thioethanol can act as a precursor for the intermediates, CH$_3$CHSH, CH$_3$CH(S)SH radical and product thioethanal. 
\begin{figure*}[t]
\centering
\begin{minipage}{0.47\textwidth}
  \centering
  \includegraphics[width=\linewidth]{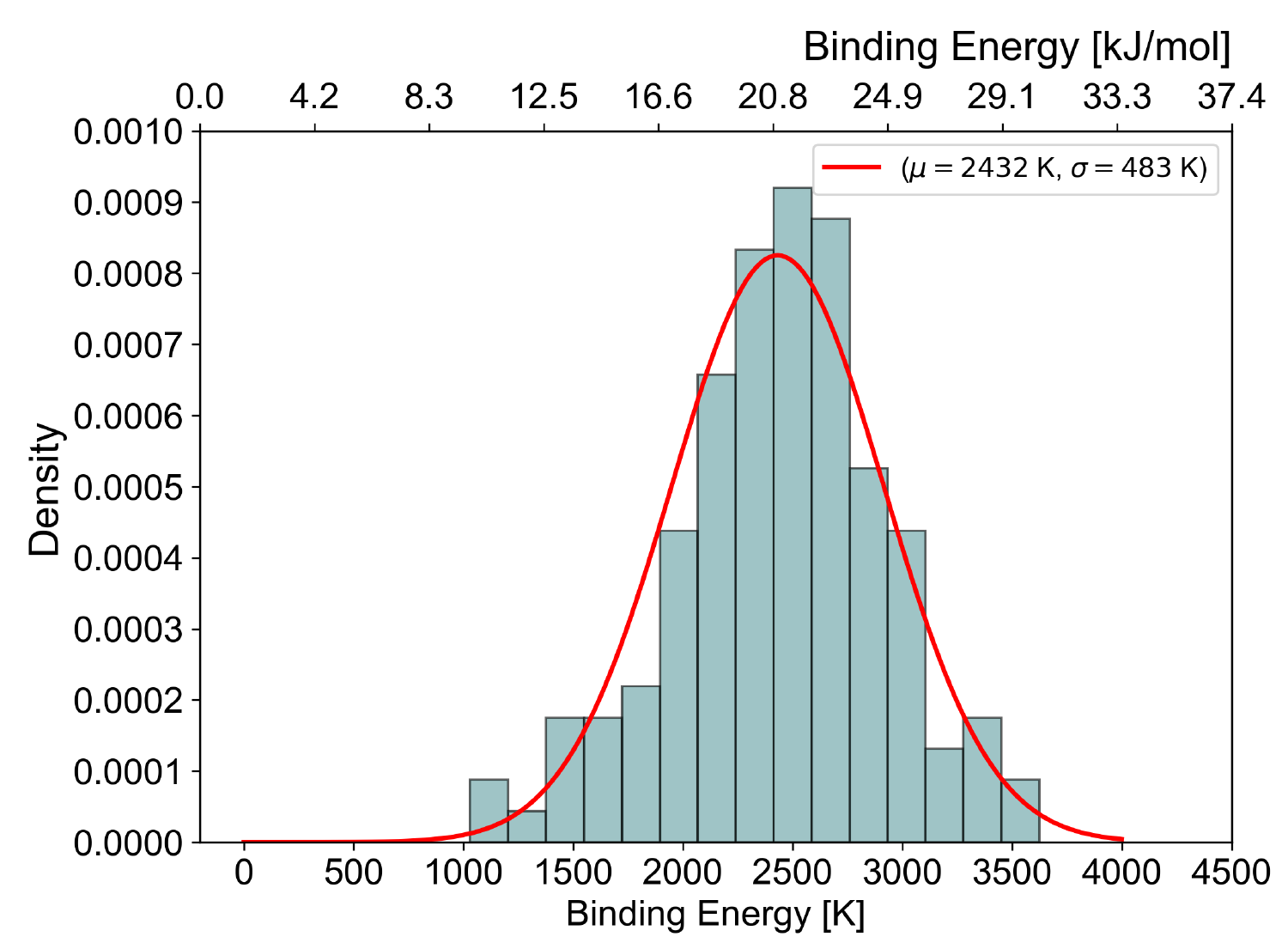}
  \caption*{(a)}
\end{minipage}
\hfill
\begin{minipage}{0.52\textwidth}
  \centering
  \includegraphics[width=\linewidth]{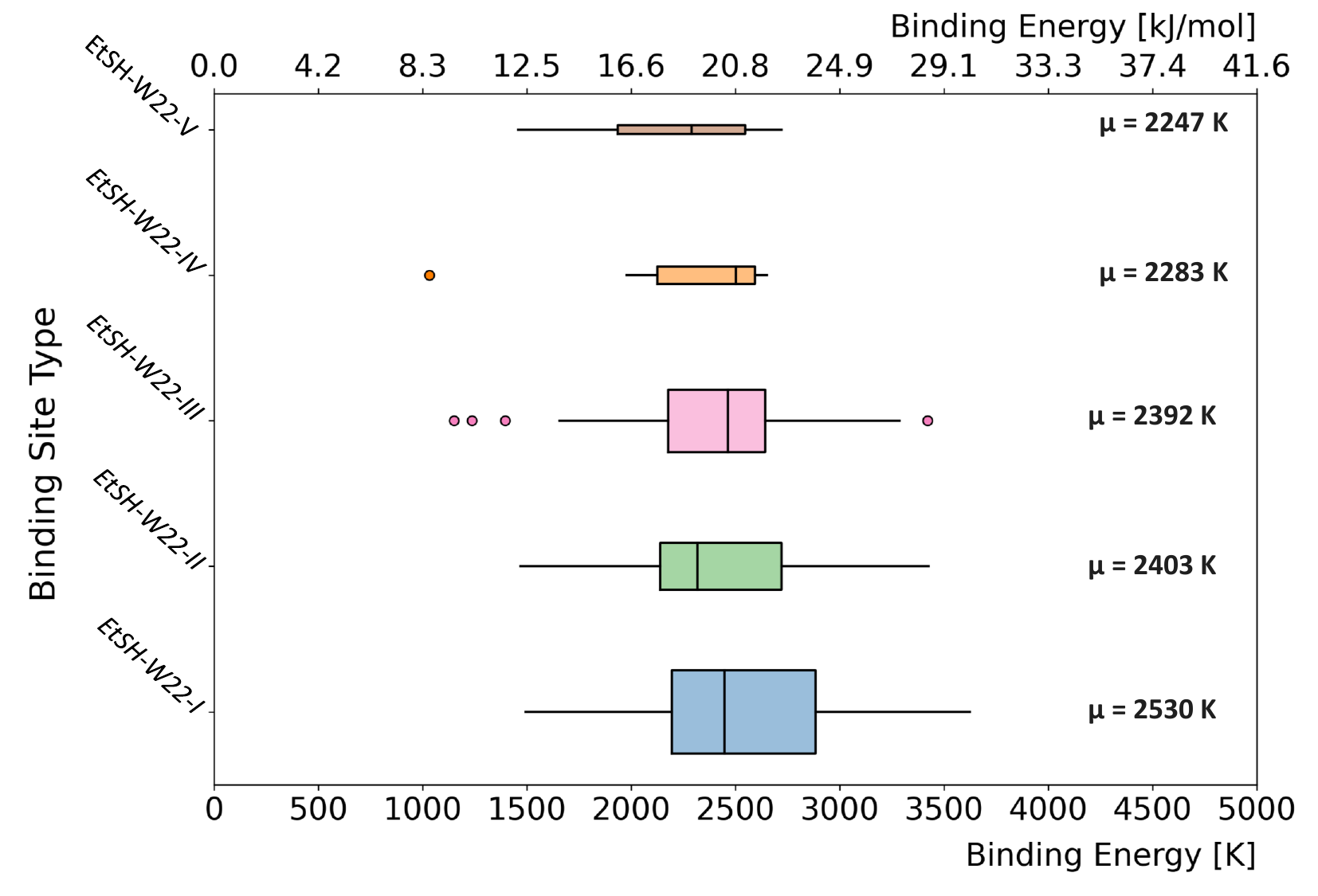}
  \caption*{(b)}
\end{minipage}
\caption{(a) Binding energy distribution of CH$_{3}$CH$_{2}$SH on the W22 ice cluster surface, illustrating the spread of binding energy values observed. (b) Box plot of binding energies for CH$_{3}$CH$_{2}$SH on different types of binding modes, showing variability and distribution across distinct binding site categories depicted in Fig.~\ref{binding sites}. \textcolor{black}{The area of each box is proportional to the percentage of binding sites in that category.}}
\label{fig:histogram_boxplot}
\end{figure*}

\begin{figure*}[tbp]
    \centering
    \includegraphics[width=\textwidth]{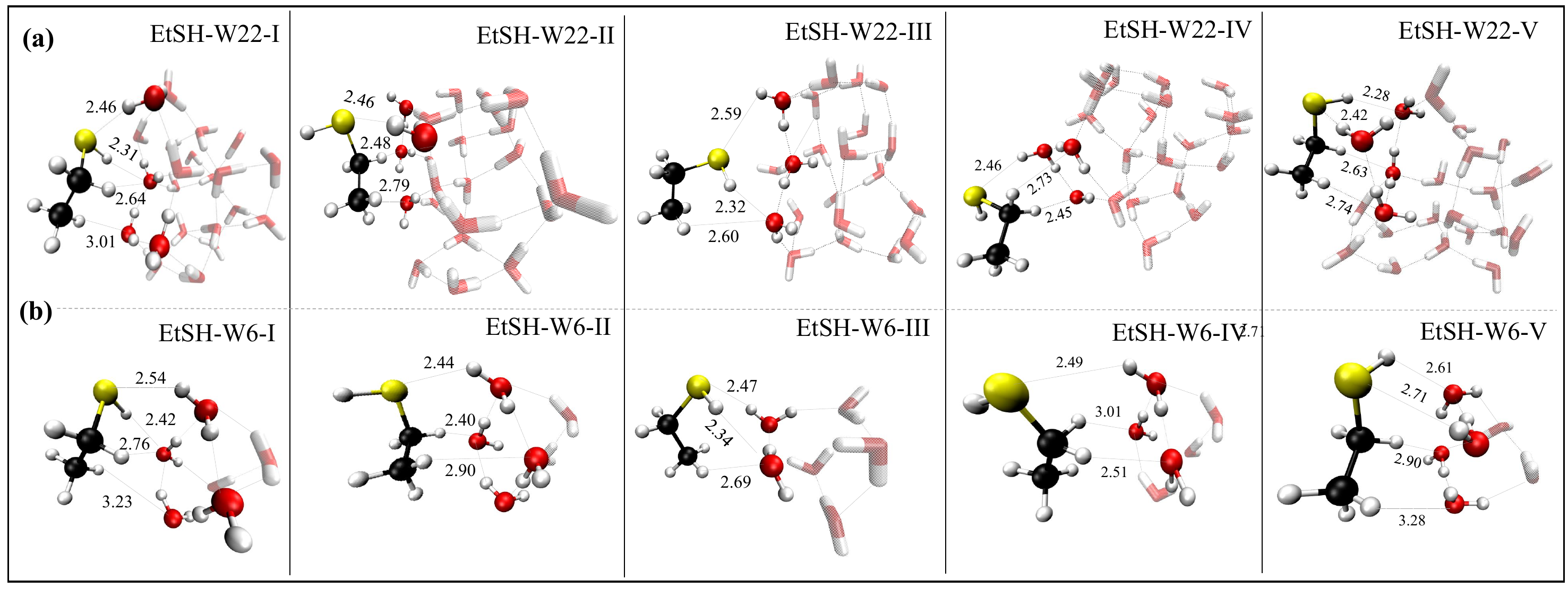}\\
    \caption{(a) Types of binding modes observed on the W22 ice cluster surface and (b) corresponding binding modes on the W6 ice cluster. The binding interactions shown illustrate how adsorbate molecules adhere to different surface sites on both clusters. Key interatomic distances are shown (in \AA), illustrating that W6 preserves the local orientation of the reactive groups. EtSH-I is used for modeling the ~\textcolor{black}{LH-initiated pathways}, while EtSH-II and EtSH-III are used for the ~\textcolor{black}{ER-initiated pathways} (details in the section \ref{Selection of Reactive Binding Sites}).}
    \label{binding sites}

\end{figure*}

\subsection{\textcolor{black}{Binding Sites of CH$_3$CH$_2$SH on ASW Clusters}} \label{subsec:Binding Sites}
A total of 130 potential binding sites for \textit{cis}-CH$_3$CH$_2$SH over the ASW cluster, W22, have been identified, exhibiting an average binding energy of 2430 K (20.2 kJ/mol). The overall binding energy distribution and the different binding site contributions are depicted in Fig.~\ref{fig:histogram_boxplot}. These binding sites are classified into five major categories based on the nature of hydrogen bonding and long-range interactions involved, as illustrated in Fig.~\ref{binding sites}. (a) The first category, EtSH-W22-I, contributes 38\% of all the binding sites. \textcolor{black}{The sulfur atom simultaneously acts as both a hydrogen-bond donor and acceptor toward the ASW surface and the molecule is bound through the -SH moiety in its \textit{cis}-configuration w.r.t the cluster}. The average binding energy value of this type is 2530 K. (b) The second category, EtSH-W22-II is the opposite of the first one with sulfur in proton acceptor interactions \textcolor{black}{and the –SH group points away from the cluster.} It contributes 21\% of the total binding sites and the average value of binding energy is 2403 K. (c) The third category, \textcolor{black}{EtSH-W22-III also, the sulfur atom simultaneously acts as both a hydrogen-bond donor and acceptor, similar to the first category,} the two methylene hydrogen atoms are in the gas phase and methyl hydrogens interacting with the cluster. This type contributes the 27\%, potential binding sites. The average value computed for this dataset is 2392 K. (d) In the fourth category (EtSH-W22-IV) which contributes 9\% of the binding sites, both hydrogen atoms interact with the cluster and are not available for abstraction by the OH radical; therefore, this binding mode is not further considered. The average BE for this binding
mode is, 2283 K. (e) The last category, EtSH-W22-V, involves the other weak binding site exhibiting long range interactions of \textit{cis}-CH$_3$CH$_2$SH and it also involves the strong binding site corresponding to the \textbf{anti}-CH$_3$CH$_2$SH, resulting into the average binding energy value of 2247 K, Fig.~\ref{binding sites}. Moreover, it contributes only 5\% of the total binding sites. Further, we located the similar binding sites by sampling \textit{cis}-CH$_3$CH$_2$SH around medium size ASW cluster containing six water molecules (W6) for computationally efficient approach to study the ice-dust grains mechanism. The key binding sites on W22 and equivalent conformations on W6 are also depicted in Fig.~\ref{binding sites}.\\
\textcolor{black}{Notably, across the W22/W6 clusters, S--H$\cdots$O (donor type) distances are of 2.28--2.61~\AA{} and O--H$\cdots$S (acceptor type) distances of 2.42--2.59~\AA{} are within the \textit{moderate} H-bond regime on ASW. In contrast, C--H$\cdots$O contacts from CH$_2$ \mbox{(2.40--3.01~\AA{})} are \textit{moderate--to--weak}, while those from CH$_3$ \mbox{(2.60--3.28~\AA{})} are predominantly \textit{weak/dispersion-assisted}. Thus, -SH centered hydrogen bonding is the primary anchoring motif, with C--H interactions providing auxiliary stabilization.
Of note, the strongest binding mode corresponds to Type~I, with the highest BE of 2530 K. As expected from the chemical versatility of sulfur, the overall distribution of binding sites is broad ($\sigma$ = 485 K). However, the mean values of the five categories differ by only $\sim$250 K, which is consistent with the dispersion-dominated behavior reported for other S-bearing species on ASW cluster \citep{perrero2022binding, bariosco2024binding}.}
\subsection{\textcolor{black}{Selection of Reactive Binding Sites}} \label{Selection of Reactive Binding Sites}
\textcolor{black}{The Type I category (EtSH-I) is selected for the ~\textcolor{black}{LH-initiated pathways} (Fig.~\ref{binding sites}) since for this mechanism both CH$_3$CH$_2$SH and OH interact directly with the surface and the CH$_2$(–H) to be abstracted should directly points towards the cluster. 
For the ER-initiated pathways, the Type II and Type III sites (EtSH-II and EtSH-III) are selected, where the required CH$_2$(–H) is into the gas phase, allowing reaction by the incoming OH radical also from the gas-phase. 
The highest BE configuration [3624 K (W22) and 2697 K (W6) for Type I, 3425 K (W22) and 2676 K (W6) for Type II, and 3421 K (W22) and 2597 K (W6) for Type III] within the each category is chosen. The corresponding W6 analogues shown in Fig.~\ref{binding sites} reproduce the steric orientation of the reactive groups, even though the absolute binding energies are systematically lower than those of W22, as expected from the smaller cluster size. However, the comparison of binding energies and key interatomic distances in Fig.~\ref{binding sites} indicates that the selected W6 analogues capture the main interaction features of the corresponding W22 sites, supporting their use as reduced models in the reactivity calculations.}

\begin{figure*}[tbp]
    \centering
    \includegraphics[width=0.95\textwidth]{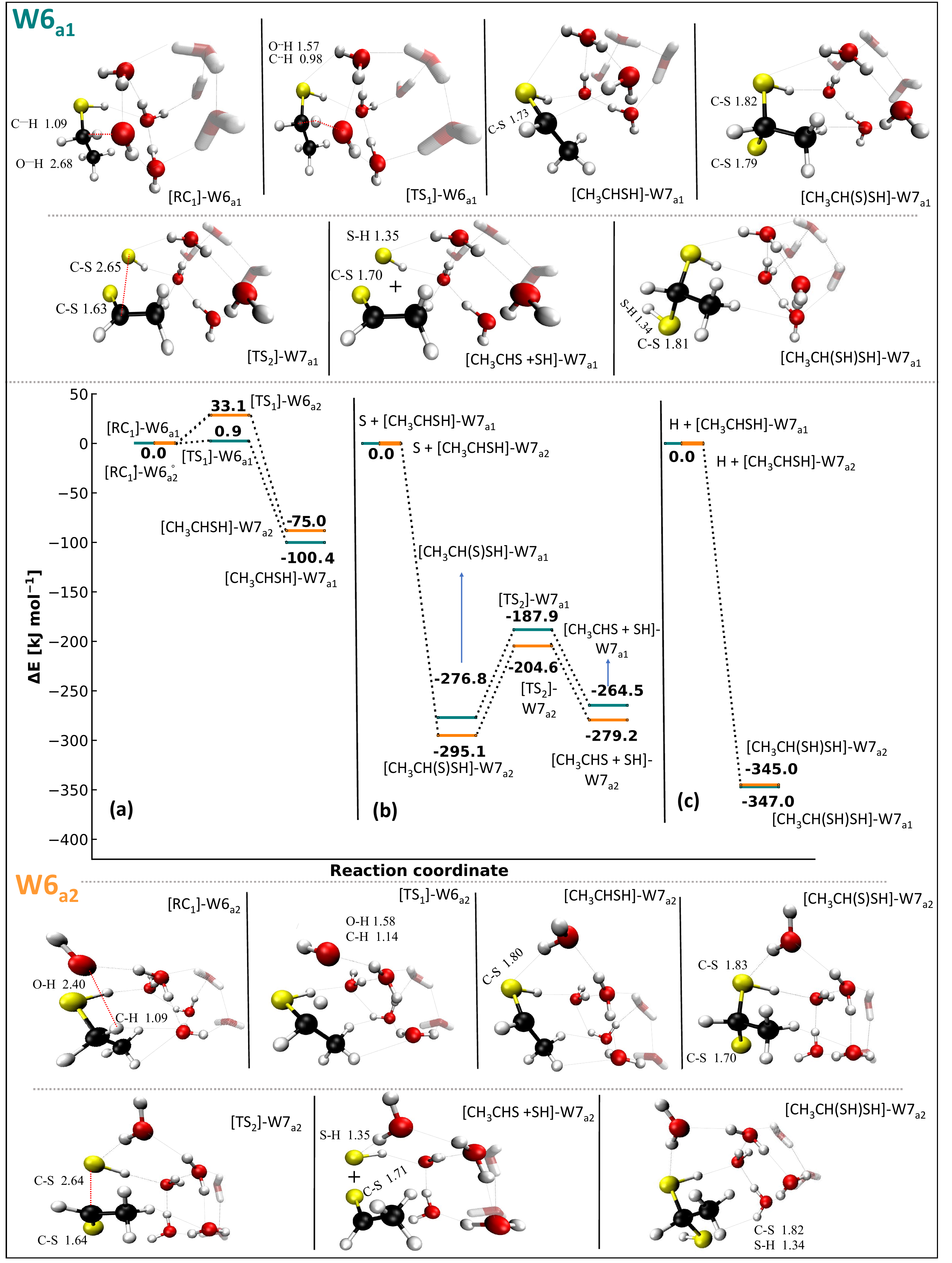}\\
    \caption{Reaction pathways, W6$_{a1}$ and W6$_{a2}$\textcolor{black}{ from the binding mode EtSH-W6-I. For (a) Step 1 and (b) Step 2a, 2b and (c) Step (c). ~\textcolor{black}{The reactants are co-adsorbed for (a) Step 1.} Geometries are optimized at the M06-2X/def2-TZVP  and the energies are refined at MPWB1K-D3(BJ)/def2-TZVPD including zero-point energy (ZPVE) corrections at the optimization level of theory.} Significant bond lengths are shown in \AA{}. The bond length involved in making and breaking are depicted in red dotted lines}.
    \label{fig:W6_a}
    
\end{figure*}
\begin{figure*}[tbp]
    \centering
    \includegraphics[width=0.95\textwidth]{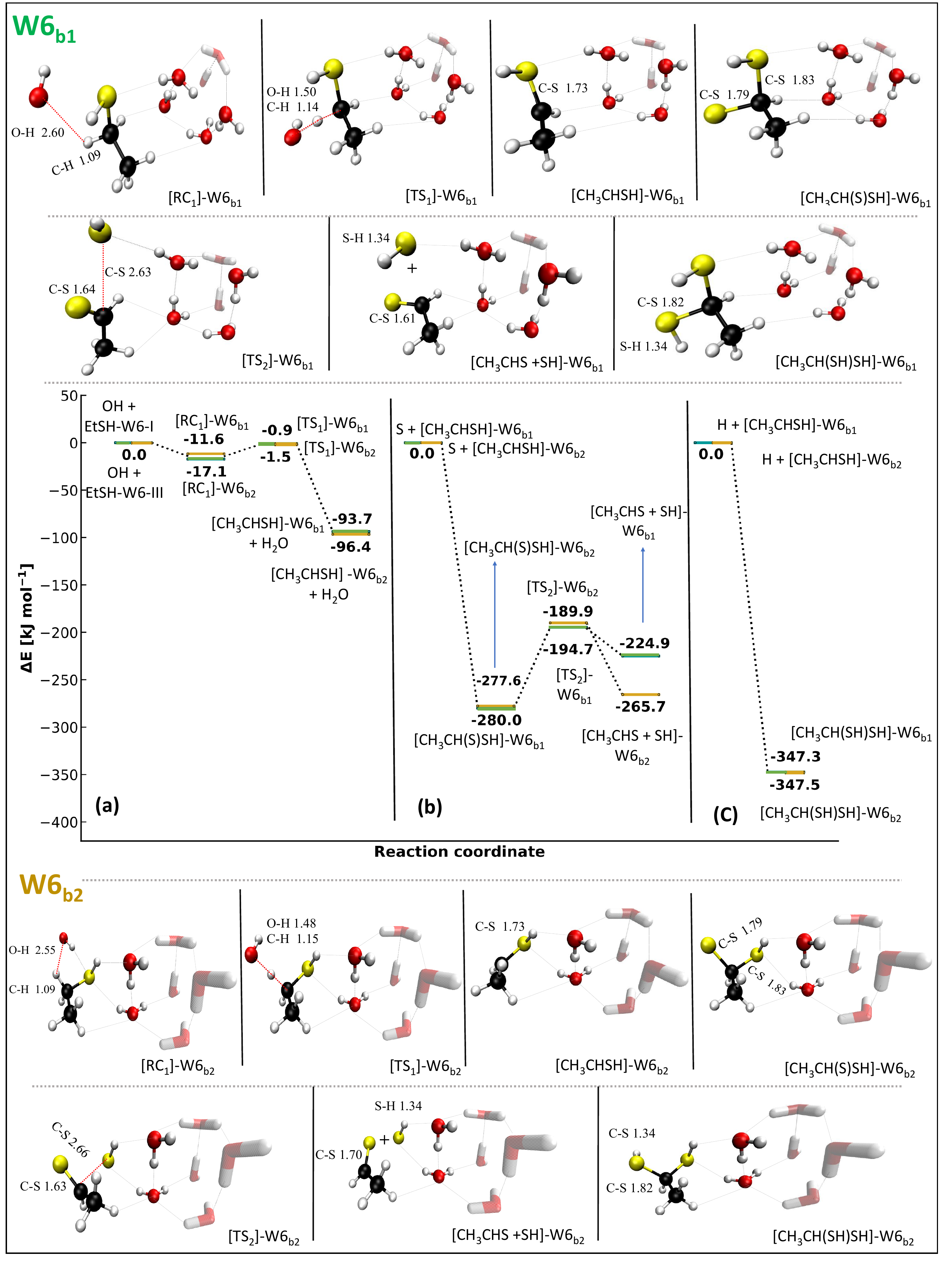}\\
    \caption{\textcolor{black}{Same as Fig.~\ref{fig:W6_a} for reaction pathway, W6$_{b1}$ and W6$_{b2}$ starting from configuration EtSH-W6-II and EtSH-W6-III, respectively. ~\textcolor{black}{CH$_3$CH$_2$SH is adsorbed on ASW and OH is in the gas-phase for (a) step 1.}}}
    \label{fig:W6_b}
    
\end{figure*}

\begin{figure*}[tbp]
    \centering
     \includegraphics[width=0.95\textwidth]{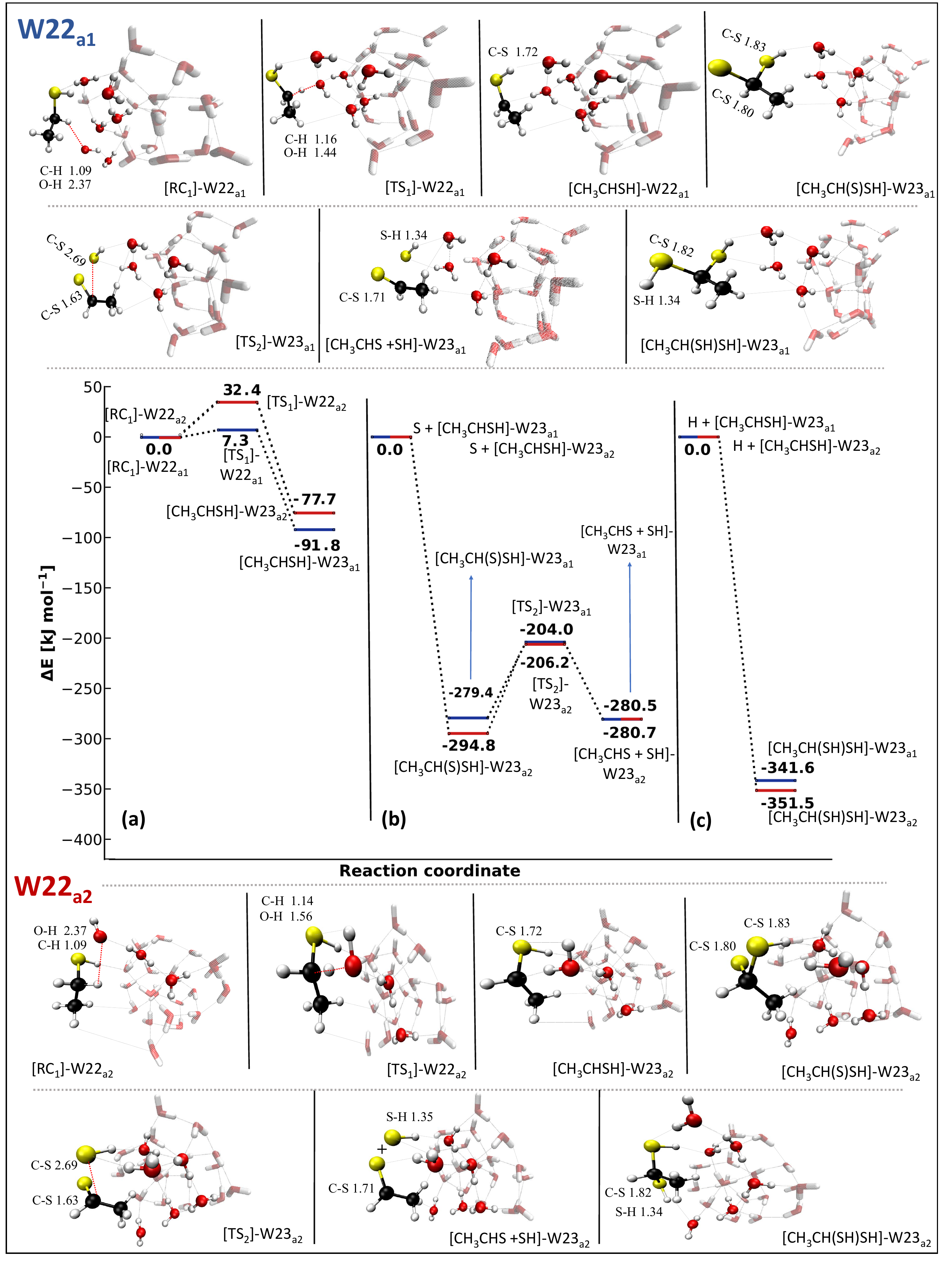}\\
    \caption{\textcolor{black}{Same as Fig.~\ref{fig:W6_a} for reaction pathway, W22$_{a1}$ and W22$_{a2}$ starting from configuration EtSH-W22-I.}}
    \label{fig:W22}
\end{figure*}

\subsection{ASW: Amorphous Solid Water Reaction Mechanism} \label{subsec:Amorphous Solid Water Reaction Mechanism}
To investigate the conversion of thioethanol CH$_3$CH$_2$SH to thioethanal (CH$_3$CHS) on an ice-dust grain, we began with W6 ASW ice cluster model to gain insights into reaction feasibility, activation barriers, intermediate stability and exothermicity of the reaction,~\textcolor{black}{depicted in Fig.~\ref{fig:W6_a} and Fig.~\ref{fig:W6_b}. The important W6 mechanisms are later extended to W22 ASW cluster, Fig.~\ref{fig:W22}. A comparative analysis of the activation energies and reaction exothermicities across different environments is presented in Table ~\ref{tab:reaction_energetics}. 
~\textcolor{black}{The results of the surface mechanisms initiated from two different pathways are discussed below.}}\\
(a) ~\textcolor{black}{Langmuir-Hinshelwood (LH) initiated pathways: 
The pathways, W6$_{a1}$ and W6$_{a2}$, derived from binding mode EtSH-W6-I depcited in Fig.~\ref{fig:W6_a} are LH-initiated pathways and involve both thioethanol and the OH radical simultaneously adsorbed forming [RC$_1$]-W6$_{a1}$ and [RC$_1$]-W6$_{a2}$, respectively}. The pathways differ in the nature and extent of the OH-cluster interaction, leading to markedly different reactivities and energetics.
In the reaction complex [RC$_1$]-W6$_{a1}$, the OH radical is fully integrated into the hydrogen bond network of the ice cluster, participating in donor and acceptor type interactions. The transition state for hydrogen abstraction, [TS$_1$]-W6$_{a1}$, is just \textcolor{black}{0.9 kJ/mol above [RC$_1$]-W6$_{a1}$}. 
\textcolor{black}{The lower activation barrier reflects} the \textcolor{black}{site-specific} catalytic role of the ice matrix in pre-orienting the reactants and increasing OH reactivity by significantly enhancing the proton affinity of the OH radical, practically removing the barrier to the H abstraction reaction. The resulting product, [CH$_3$CHSH]-W7$_{a1}$, is further stabilized by reintegration of H$_2$O into the cluster, ~\textcolor{black}{with an overall exothermicity of 100.4 kJ/mol.}\\
By contrast, in the [RC$_1$]-W6$_{a2}$, the OH radical interacts only weakly with the cluster through a single donor-type hydrogen bond and is not embedded in the H-bond ice-matrix. The hydrogen abstraction barrier is increased to \textcolor{black}{33 kJ/mol}. The limited interaction of [RC$_1$]-W6$_{a2}$ with the surface prevents pre-orientation into an optimal reactive geometry. As a result, the transition state is less stabilized, leading to a higher activation barrier. Unlike W6$_{a1}$, the cluster in W6$_{a2}$ does not contribute significantly to OH activation and instead imposes geometric constraints that increases the energy of the transition state.\\
In both the pathways W6$_{a1}$ and W6$_{a2}$, the subsequent radical–radical addition of atomic sulfur to [CH$_3$CHSH]-W7, forms radical intermediate [CH$_3$CH(S)SH]-W7 and is highly exothermic, 275.8 kJ/mol and 295.1 kJ/mol respectively for W6$_{a1}$ and W6$_{a2}$, Fig.~\ref{fig:W6_a}(b). The dissociation of this intermediate to [CH$_3$CHS + SH]-W7 proceeds via a transition state [TS$_2$]-W7, with similar barrier for W6$_{a1}$ and W6$_{a2}$. The final product complex [CH$_3$CHS + SH]-W7 is significantly stabilized by 264.5 kJ/mol and 279.2 kJ/mol for W6$_{a1}$ and W6$_{a2}$ respectively. The additional exothermicity of later is attributed to strong hydrogen-bonding interactions involving the SH fragment and the ice surface in the final state, Fig.~\ref{fig:W6_a}(b). 
The alternative hydrogenation pathway involves [CH$_3$CH(S)SH]-W7 to form ethane-1,1-di-thiol [(CH$_3$CH(SH)SH)]-W7, Fig.~\ref{fig:W6_a}(c). \textcolor{black}{This step, produces a thermodynamically stable product and is significantly viable on ice-grain surfaces due to the presence of mobile H atoms and the possibility of the cluster to dissipate the excess reaction energy.}\\
(b) ~\textcolor{black}{Eley–Rideal (ER) initiated pathways}: Two different pathways, W6$_{b1}$ and W6$_{b2}$ are depicted in Fig.~\ref{fig:W6_b}. 
\textcolor{black}{In the W6$_{b1}$, derived from binding mode EtSH-W6-II, the adsorbed CH$_3$CH$_2$SH reacts with the gas-phase OH radical, Fig.~\ref{fig:W6_b}(a). The reactive complex, [RC$_1$]-W6$_{b1}$, and the transition state, [TS$_1$]-W6$_{b1}$, are equally stabilized by the interactions with the ice-cluster and the activation barrier of the initial hydrogen abstraction step is 16.2 kJ/mol with respect to [RC$_1$]-W6$_{b1}$ and exothermicity is 93.7 kJ/mol. The newly formed H$_2$O does not directly engage with the ice-matrix, unlike ~\textcolor{black}{LH-initated pathways}.\\
For the pathway, W6$_{b2}$, the reaction complex [RC$_1$]-W6$_{b2}$ is traced from the binding mode EtSH-W6-III and the hydrogen abstraction barrier is 10.1 kJ/mol with an exothermicity 96.4 kJ/mol.}\\  
The subsequent radical–radical coupling step to form the intermediate of type [CH$_3$CH(S)SH]-W6 in W6$_{b1}$ and W6$_{b2}$ is highly exothermic, Fig.~\ref{fig:W6_b}(b). The dissociation barrier associated with [TS$_2$]-W6 type transition state is 85.3 kJ/mol and 87.7 kJ/mol above the intermediate [CH$_3$CH(S)SH]-W6$_{b1}$ and [CH$_3$CH(S)SH]-W6$_{b2}$, repectively, Fig.~\ref{fig:W6_b}(b). The final product, [CH$_3$CHS + SH]-W6, formation is also highly exothermic competed by the di-thiol [CH$_3$CH(SH)SH-W6] formation by hydrogenation of [CH$_3$CH(S)SH]-W6, with an exothermicity of {\textasciitilde}347 kJ/mol for W6$_{b1}$ and W6$_{b2}$, Fig.~\ref{fig:W6_b}(c).\\
~\textcolor{black}{The key results from the W6 cluster mechanism (Table~\ref{tab:reaction_energetics}) can be summarized as follows.
(i) No substantial change in the activation barrier for the hydrogen abstraction in Step~\ref{eq:reaction1} is observed for the ~\textcolor{black}{ER-initiated pathways}, and the barrier ($\Delta E^{\ddagger}$) remains comparable to the gas-phase value. In contrast, the activation barriers along the ~\textcolor{black}{LH-initiated pathways} are modulated by the ice environment, exhibiting site-specific variations. Among these, the W6$_{a1}$ configuration emerges as the most favorable, with a significantly reduced activation barrier.
(ii) For Step~\ref{eq:reaction1}, the deviation in reaction exothermicity ($\Delta E$) on the W6 ASW cluster is generally small relative to the gas phase, while a more pronounced decrease is observed for the W6$_{a2}$. Similarly smaller deviations are found for the exothermicity of Step~\ref{eq:reaction2a} ($\Delta E_{\mathrm{a}}$) except an increase in the value for W6$_{a2}$.
(iii) The dissociation barrier on the W6 ASW cluster does not differ significantly from that obtained in the gas phase.
(iv) Compared to the gas-phase reaction, the formation of the targeted product CH$_3$CHS is more exothermic on the W6 cluster, as reflected by the larger ($\Delta E_{\mathrm{b}}$). In contrast, the formation of the secondary product CH$_3$CH(SH)SH shows no significant change in reaction exothermicity ($\Delta E_{\mathrm{c}}$) relative to the gas phase.}\\
To assess the impact of a larger and more realistic ice environment, we therefore extended our investigation of the ~\textcolor{black}{LH-initiated pathways}, which exhibit pronounced site-specific variations on W6, to a larger 22-water amorphous ice cluster (W22), traced from the binding mode EtSH-W22-I.\\
In the W22$_{a1}$ pathway, Fig.~\ref{fig:W22}(a), the OH radical is fully embedded within the ice matrix, while thioethanol is adsorbed via binding mode W22-I. The reaction complex, [RC$_1$]-W22$_{a1}$, is stabilized through extensive hydrogen bonding, and the activation barrier for hydrogen abstraction via [TS$_1$]-W22$_{a1}$ is 
7.3 kJ/mol. 
This first step is highly exothermic, due to enhanced stabilization of the product [CH$_3$CHSH + H$_2$O]-W23$_{a1}$ within the extended cluster. 
In contrast, the W22$_{a2}$, Fig.~\ref{fig:W22}, pathway features a non-embedded OH radical that interacts more weakly with the surface. in W6$_{a2}$, this results in a significantly increased activation barrier of \textcolor{black}{32.4} kJ/mol for the H-abstraction step, [TS$_1$]-W22$_{a2}$ relative to [RC$_1$]-W22$_{a2}$, and the exothermicity to form [CH$_3$CHSH + H$_2$O]-W22$_{a2}$ is  91.8 kJ/mol. 
\textcolor{black}{The subsequent exothermic sulfur addition in W22$_{a1}$ and W22$_{a2}$ forms the radical intermediate of type [CH$_3$CH(S)SH]-W23. The intermediate dissociates into the targeted product [CH$_3$CHS + H$_2$O]-W23 via [TS$_2$]-W23$_{a1}$ or [TS$_2$]-W23$_{a2}$ assoicated with barrier of 75.4 kJ/mol or 88.6 kJ/mol respectively for W22$_{a1}$ and W22$_{a2}$. The former dissociation barrier is lower due to the additional stabilization of the associated transition state. The final product is further stabilized by additional
hydrogen-bonding interactions on to the larger cluster with a release of {\textasciitilde}280 kJ/mol for W22$_{a1}$ and W22$_{a2}$. Alternatively, the hydrogenation of [CH$_3$CH(S)SH + H$_2$O]-W23 can also form di-thiol secondary product [CH$_3$CH(SH)SH + H$_2$O]-W23, Fig.~\ref{fig:W22}.\\
Analyzing, the site-specific variation of activation barrier for Step ~\ref{eq:reaction1} on the W22 cluster with respect the gas-phase mechanism, Table ~\ref{tab:reaction_energetics},  it is noted that the barrier decreased in W22$_{a1}$ and significantly increased for W22$_{a2}$, similar to that observed for W6 cluster. Notably, the magnitude of the reduction in W22$_{a1}$ is comparable to the estimated uncertainty of the employed level of theory in the gas-phase (Appendix ~\ref{appendix:DFT}, ~\ref{fig:bnchmark_rxnEnergy}), But, in W22$_{a1}$ the barrier is unlikely to exceed the gas-phase value by significant amount, even within the associated error bars. Further, the exothermicity of the first step, $\Delta E$, on the W22 cluster is reduced relative to the gas phase similar to the W6 cluster. For Step 2 also, W22 shows a similar trend to that observed for the W6.}
Overall, the effect of the larger water cluster, W22, is relatively modest compared to W6, suggesting that the six-water model already captures the key trends in energetics and reactivity. While, W22, provides enhanced stabilization of some final products but the reaction barriers and thermodynamic profiles remain largely consistent. \textcolor{black}{However, neither the W6 nor the W22 cluster accounts for extended hydrogen-bonding networks or cavity-like surface morphologies. As discussed in \cite{bovolenta2020high} and \cite{ferrero2020binding}, it can lead to enhanced binding and altered reactivity compared to flat surfaces. \cite{martinez2023gas} further highlight that cavity-mediated reactivity differs significantly from crystalline models.}
\begin{table}[tbp]
\centering
\small
\setlength{\tabcolsep}{4pt}
\renewcommand{\arraystretch}{1.12}

\centering
\caption{\textcolor{black}{Reaction energetics for CH$_3$CHS in different environment for Step \ref{eq:reaction1}, Step \ref{eq:reaction2a}, \ref{eq:reaction2b} and \ref{eq:reaction2c}; and comparison with CH$_3$CHO synthesis in the gas-phase. Energies are given in kJ/mol.}}
\label{tab:reaction_energetics}
\begin{tabular}{|c|cc|cccc|}
    \hline
    \textbf{} & \multicolumn{2}{|c|}{\textbf{STEP 1}} & \multicolumn{4}{|c|}{\textbf{STEP 2}} \\ \hline
    \textbf{} & $\Delta E^{\ddagger}$ & $\Delta E$ 
& $\Delta E_{\mathrm{a}}$
& $\Delta E_{\mathrm{b}}^{\ddagger}$ 
& $\Delta E_{\mathrm{b}}$ 
& $\Delta E_{\mathrm{c}}$ \\ \hline
Gas-Phase & \textcolor{black}{10.9} & \textcolor{black}{-102.7} & \textcolor{black}{-275.9} & \textcolor{black}{87.1} & \textcolor{black}{-195.0} & \textcolor{black}{-352.5} \\
W6$_{a1}$ & 0.9 & \textcolor{black}{-100.4} & -276.8 & 88.9 & -264.5 & -347.0 \\
W6$_{a2}$ & \textcolor{black}{33.0} & \textcolor{black}{-75.0} & \textcolor{black}{-295.1} & \textcolor{black}{90.5} & \textcolor{black}{-279.2} & \textcolor{black}{-345.0} \\ 
W6$_{b1}$ & \textcolor{black}{16.2} & \textcolor{black}{-93.7} & \textcolor{black}{-280.0} & \textcolor{black}{85.3} & \textcolor{black}{-224.9} & \textcolor{black}{-347.3} \\ 
W6$_{b2}$ & \textcolor{black}{10.1} & \textcolor{black}{-96.4} & \textcolor{black}{-277.6} & \textcolor{black}{87.7} & \textcolor{black}{-265.7} & \textcolor{black}{-347.5} \\ 
W22$_{a1}$ & \textcolor{black}{7.3} & \textcolor{black}{-77.7} & \textcolor{black}{-279.4} & \textcolor{black}{75.4} & \textcolor{black}{-280.7} & \textcolor{black}{-341.6} \\ 
W22$_{a2}$ & \textcolor{black}{32.4} & \textcolor{black}{- 91.8} & \textcolor{black}{-294.8} & \textcolor{black}{88.6} & \textcolor{black}{-280.5} & \textcolor{black}{-351.5} \\ \hline
    \multicolumn{7}{|c|}{\textbf{For CH$_3$CHO synthesis \cite{martinez2023gas}}} \\ \hline
    Gas-phase & 18.1 & -95.0 & -292.1 & 72.8 & -312.0 & -- \\ \hline

\bottomrule
\end{tabular}

\tablefoot{Step~1: $\Delta E^{\ddagger}=(\mathrm{TS}_1)-(\mathrm{RC}_1)$ and
$\Delta E=(\mathrm{CH}_3\mathrm{CHSH}+\mathrm{H}_2\mathrm{O})-(\mathrm{CH}_3\mathrm{CH}_2\mathrm{SH}+\mathrm{OH})$; \textcolor{black}{but for (W6/W22)$_{a1/a2}$, $\Delta E=(\mathrm{CH}_3\mathrm{CHSH}+\mathrm{H}_2\mathrm{O})-(\mathrm{RC}_1)$.}
Step~2: $\Delta E_{\mathrm{a}}=(\mathrm{CH}_3\mathrm{CH(S)SH})-(\mathrm{CH}_3\mathrm{CHSH}+\mathrm{S})$,
$\Delta E_{\mathrm{b}}^{\ddagger}=(\mathrm{TS}_2)-(\mathrm{CH}_3\mathrm{CH(S)SH})$,
$\Delta E_{\mathrm{b}}=(\mathrm{CH}_3\mathrm{CHS}+\mathrm{SH})-(\mathrm{CH}_3\mathrm{CH(S)H}+\mathrm{S})$,
$\Delta E_{\mathrm{c}}=(\mathrm{CH}_3\mathrm{CH(SH)SH})-(\mathrm{CH}_3\mathrm{CH(S)SH}+\mathrm{H})$.
For W6/W22, CH$_3$CH$_2$SH as well as other intermediates/transition states are adsorbed \textcolor{black}{and (S/H) are in the gas phase. OH is also in the gas-phase for W6$_{b1/b2}$}}
\end{table}

\section{ Discussion} \label{sec:Discussion}
Our results demonstrate that the conversion of thioethanol (CH$_3$CH$_2$SH) to thioethanal (CH$_3$CHS) proceeds via an overall barrierless mechanism, viable under the low-temperature conditions of dense molecular clouds, \textcolor{black}{in the gas-phase.} The rate-determining step is the hydrogen abstraction forming CH$_3$CHSH, while the subsequent addition of atomic sulfur is highly exothermic and effectively compensates for the moderate dissociation barrier of the intermediate CH$_3$CH(S)SH. The fate of this stable intermediate diverges depending on the environment; In the gas phase, inefficient energy dissipation favors prompt dissociation to CH$_3$CHS, whereas on ice surfaces, energy is rapidly quenched into the substrate, promoting kinetic trapping. In hydrogen-rich regions, CH$_3$CH(S)SH may instead undergo \textcolor{black}{highly exothermic} hydrogenation to form ethane-1,1-di-thiol (CH$_3$CH(SH)SH), a thermodynamically preferred side product. Thus, CH$_3$CHS formation competes with both intermediate stabilization and hydrogenation pathways, \textcolor{black}{in addition to the high-energy transition state of step ~\ref{eq:reaction2b}}, with the dominant outcome dictated by local temperature, density, and whether the chemistry occurs in the gas phase or on icy grains. Notably, our computations indicate that \textit{geminal} ethane-1,1-di-thiol is more likely to form from thioethanol than the \textit{vicinal} isomer observed in recent ice experiments by \cite{santos2024formation}. This aligns with studies on oxygen analogues, where ethan-1,1-diol is the global minimum structure on the C$_2$H$_6$O$_2$ potential energy surface ~\citep{noriega2024quest}. We therefore propose that ethane-1,1-di-thiol should be considered a promising candidate for future astrochemical detection and modeling.\\
A central mechanistic insight from this work is the critical role of the ice-surface structure in modulating reactivity. Unlike a uniform catalytic effect, surface contributions are highly dependent on adsorption geometry and hydrogen-bond topology. When the OH radical is fully embedded in the H-bond matrix, as in W6$_{a1}$ and W22$_{a1}$, the abstraction barrier is reduced due to enhanced proton affinity and pre-orientation, reflecting the \textcolor{black}{site-specific} catalytic potential of amorphous ice surfaces. In contrast, when OH is only weakly or partially adsorbed (W6$_{a2}$, W22$_{a2}$), the lack of stabilizing hydrogen bonds and poor alignment leads to significantly higher barriers, in some cases even exceeding those in the gas phase. The~\textcolor{black}{Eley–Rideal initiated pathways}, where OH remains in the gas phase, show minimal surface influence unless both reactants engage directly with the ice.
\textcolor{black}{
It emphasize that interstellar ices can either lower or increase reaction barriers, depending on local adsorption geometry, which highlights the importance of detailed binding-site sampling for accurate reactivity predictions}.\\
When compared with its oxygen analogues, sulfur chemistry shows clear mechanistic deviations. Prior studies on the hydrogen abstraction from the ethanol \citep{ocana2018gas, carr2011site, xu2007theoretical, caravan2015measurements, zheng2012multi} report different abstraction preferences than observed for thioethanol in this study, Fig.~\ref{fig:step1}. Due to the weaker -S-H bond strength compared to -O-H the CH$_3$CH$_2$S formation has a more favorable energetic profile than its oxygen counterpart, CH$_3$CH$_2$O.
 This introduces a competing pathway in the sulfur case, where a significant portion of thioethanol may be converted into the CH$_3$CH$_2$S radical instead of progressing toward CH$_3$CHS formation. As a result, less CH$_3$CHS is expected compared to its oxygen analogue, ethanal. Moreover, previous astrochemical modeling studies ~\citep{agundez2025detection} have shown that CH$_3$CH$_2$S does not efficiently convert into CH$_3$CHS, further supporting the idea that CH$_3$CHS formation is less favorable from CH$_3$CH$_2$S. These factors together may explain the lower abundance and delayed detection of CH$_3$CHS in the interstellar medium.\\
 Moreover, in comparison with the analogous oxygen-based mechanism studied by \cite{martinez2023gas} \textcolor{black}{in the gas-phase, key} differences are observed from Table~\ref{tab:reaction_energetics}. First, the thioethyl radical intermediate, CH$_3$CHSH, is more stable than its oxygen counterpart, CH$_3$CHOH, whereas the product thioethanal, CH$_3$CHS, is thermodynamically less stable than ethanal, CH$_3$CHO. This could be another possible reason for the non-detection of CH$_3$CHS in the ISM in varied astrophysical environment unlike CH$_3$CHO. However, the intermediates located in the examined mechanism, CH$_3$CHSH and CH$_3$CH(S)SH, might be stable enough to be observed. Further insights will be obtained from a comparative study of the binding energy distributions of these species to derive an estimate of the intermediate and products lifetime on ASW, which will be the focus of our forthcoming studies.

\section{Astrophysical Implications} \label{sec:cite}
The reaction pathways investigated in this work provides an efficient gas-phase formation route for CH$_3$CHS from the CH$_3$CH$_2$SH via dissociation of radical CH$_3$CH(S)SH and the addition of atomic hydrogen to this radical forms the ethane-1,1-di-thiol, CH$_3$CH(SH)SH as a competing channel on the ice surfaces. So, in cold, quiescent regions like TMC-1, where, the abundance of atomic hydrogen is low because most hydrogen exists in the molecular form (H$_2$), and the region is well shielded from intense UV radiation, the reaction pathways leading to the di-thiol formation~\textcolor{black}{may be suppressed, which could favor the persistence of CH$_3$CHS in the gas-phase.} In contrast, in warmer and more active star-forming regions such as Orion, strong UV radiation from young massive stars and the presence of HII regions are \textcolor{black}{expected to increase} the abundance of atomic hydrogen. This environment may allow multiple competing sulfur-bearing reaction pathways to operate on ice-grain surfaces, potentially enabling the formation of both CH$_3$CHS and di-thiol formation. \textcolor{black}{In highly active environments such as Sgr B2, where dense molecular gas coexists with intense UV radiation and frequent energetic events, elevated atomic hydrogen abundances may further enhance competing reaction channels, which could reduce the relative gas-phase abundance of CH$_3$CHS}.\\
Further, our computed binding energy of CH$_3$CH$_2$SH (thioethanol) on interstellar ice surfaces, 2430 K, provides the first reliable value for this molecule, filling a gap in existing astrochemical data. The availability of an accurate value is essential for modeling sulfur chemistry on icy grains, especially in cold interstellar environments where desorption rates and surface residence times are highly sensitive to binding energies.\\
In comparison, \textcolor{black}{ethanol (CH$_3$CH$_2$OH) has a higher binding energy (3127 K - 7108 K) \citep{perrero2024binding}, making it significantly less volatile.} 
\textcolor{black}{Despite this, ethanol is detected at much higher abundances in comparison to thioethanol, in regions such as Orion KL and Sgr B2(N2),\citep{kolesnikova2014spectroscopic, muller2016exploring, martin2019submillimeter}, possibly due to the higher cosmic abundance of oxygen compared to sulfur \citep{jenkins2009unified} }.\\
Additionally, while ethanol \textcolor{black}{is likely to} remain predominantly sequestered in icy mantles at low temperatures due to its high binding energy, thioethanol’s greater volatility and broader binding energy distribution suggest that sulfur-bearing organics \textcolor{black}{may be less prone to long-term retention on ice surfaces.} This dynamic behavior \textcolor{black}{could} facilitate more frequent exchange between solid and gas phases and increases the likelihood of chemical processing and destruction once sulfur species enter the gas phase. Such reactivity \textcolor{black}{might also} contribute to comparatively low steady-state abundances of specific sulfur-bearing organic molecule, despite episodic desorption events. \textcolor{black}{These qualitative considerations highlight the need for dedicated astrochemical modeling to assess their quantitative impact over broader discussion of sulfur depletion in the ISM.}\\
Moreover, in warmer star-forming regions like Orion and Sgr B2, thermal desorption becomes effective, enabling CH$_3$CH$_2$SH to desorb, ~\textcolor{black}{likely} from both low- and high-energy binding sites, which supports its sustained presence in the gas phase and subsequent detection \citep{kolesnikova2014spectroscopic, muller2016exploring, martin2019submillimeter}.\\
However, it is important to note that under the extreme cold conditions of TMC-1 (\textasciitilde10 K), thermal desorption is highly inefficient, even from low binding energy sites. Therefore, any release of CH$_3$CH$_2$SH into the gas phase \textcolor{black}{is expected to} occur via non-thermal desorption mechanisms such as cosmic-ray induced heating, reactive desorption, or UV photodesorption. The latter is particularly relevant in the outer envelope or more diffuse peripheral regions of TMC-1, where the visual extinction (A\textsubscript{V}) is lower and interstellar UV radiation can penetrate more effectively, enhancing UV photodissociation and photodesorption processes \citep{fuente2019gas}. \textcolor{black}{Once desorbed CH$_3$CH$_2$SH is expected to undergo barrierless thermal conversion to CH$_3$CHS as proposed in this work and photochemical conversion under UV irradiation, as demonstrated by laboratory experiments \citep{purzycka2021uv}.}
\\
\textcolor{black}{This may indicate that the dominant desorption mechanisms, non-thermal in cold environments and thermal in warmer star-forming regions, could affect the detectability and chemical evolution of sulfur-bearing species in the interstellar medium. The tentative hpothesis of non-detection of CH$_3$CH$_2$SH in TMC-1, contrasted with its detection in Orion and Sgr B2(N) \citep{kolesnikova2014spectroscopic,muller2016exploring,martin2019submillimeter}, together with the exclusive detection of CH$_3$CHS \citep{agundez2025detection} in TMC-1, tends to support this chemical differentiation, however, needs to be validated through astrochemical modeling.}


\newenvironment{tightitemize}{
\begin{itemize}
  \setlength{\itemsep}{0pt}
  \setlength{\parskip}{0pt}
  \setlength{\parsep}{0pt}
}{\end{itemize}}

\section{Conclusions}

\textcolor{black}{In this work, we explore the binding energy distribution, gas-phase and surface reactivity of CH$_3$CH$_2$SH under interstellar conditions, focusing on its potential conversion to CH$_3$CHS. These results provide new mechanistic insights into sulfur-bearing organic chemistry in cold dense molecular clouds. 
The main findings of this study can be summarized as follows: }

\begin{itemize}
    \item \textcolor{black}{By means of high level \textit{ab initio} quantum chemistry, we found an effectively barrierless, exothermic gas-phase formation mechanism for thioethanal (CH$_3$CHS), from thioethanol (CH$_3$CH$_2$SH). The reaction preferentially proceeds via hydrogen abstraction from the methylene group, forming the CH$_3$CHSH radical, followed by atomic sulfur addition to yield the intermediate CH$_3$CH(S)SH, which subsequently dissociates into CH$_3$CHS.}
    
    
    \item \textcolor{black}{On ice-grain surfaces, an additional competing pathway is presented, where the formation of CH$_3$CH(S)SH becomes thermodynamically favored, potentially reducing the efficiency of CH$_3$CHS formation under interstellar conditions. The fate of the intermediate CH$_3$CH(S)SH, whether it dissociates, becomes kinetically trapped, or undergoes hydrogenation to form a di-thiol, depends on the physical and chemical environment. Thus, we predict the formation of \textit{geminal} ethane-1,1-di-thiol as a possible product in hydrogen-rich environments on ice-grain surfaces, identifying it as a target for future interstellar detection.}
    \item{We show that on the cluster ice models, depending on the local hydrogen-bonding topology,
    of the OH radical, the activation barrier of the deportonation pf \ce{CH3CH2SH} can either increase or decrease. This 
    highlights the importance of considering site-specific effects rather than assuming uniform catalytic behavior.} 
    \item The study provides the first high-level binding energy estimate of CH$_3$CH$_2$SH on water ice ($\mu = 2430$K), supplying a key parameter for astrochemical modeling.
    
\end{itemize}
These results suggest potential \textcolor{black}{factors that could contribute to the low} abundance or delayed detection of CH$_3$CHS in the interstellar medium and offer \textcolor{black}{tentative hypotheses for the mutually exclusive detections of CH$_3$CHS and CH$_3$CH$_2$SH in TMC-1, Orion, and SgrB2(N); and calls for future quantitative astrochemical modeling to assess the hypotheses whether the observed trends reflect intrinsic chemical processes or observational constraints.} Consequently, \textcolor{black}{it can help inform} future observational searches with facilities such as ALMA and JWST, to target di-thiol CH$_3$CH(SH)SH along with the radical species, CH$_3$CHSH, CH$_3$CH(S)SH reported in this work.

\begin{acknowledgements}
      NR thanks FONDECYT POSTDOCTORADO (ANID) grant 3230221 for financial support and Gabriela Silva Vera for discussions and initial assistance with BEEP. SVG thanks VRID research grant 2022000507INV for financing this project. SB acknowledges BASAL Centro de Astrofisica y Tecnologias Afines (CATA), project number AFB-17002. The authors thanks Dr. Mauro Satta for fruitful discussions on the gas-phase pathways. 
\end{acknowledgements}
\section*{Data Availability}
The XYZ coordinates of all optimized structures and benchmark datasets used in this study will be deposited in a Zenodo repository and made publicly accessible upon manuscript acceptance. Additional computational details are available from the corresponding author upon request.


\bibliographystyle{aa}
\bibliography{references}

\onecolumn
\begin{appendix}

\section{DFT Benchmarking Criteria}\label{appendix:DFT}

\begin{figure}[ht]
    \centering
    \includegraphics[height=0.35\textheight]{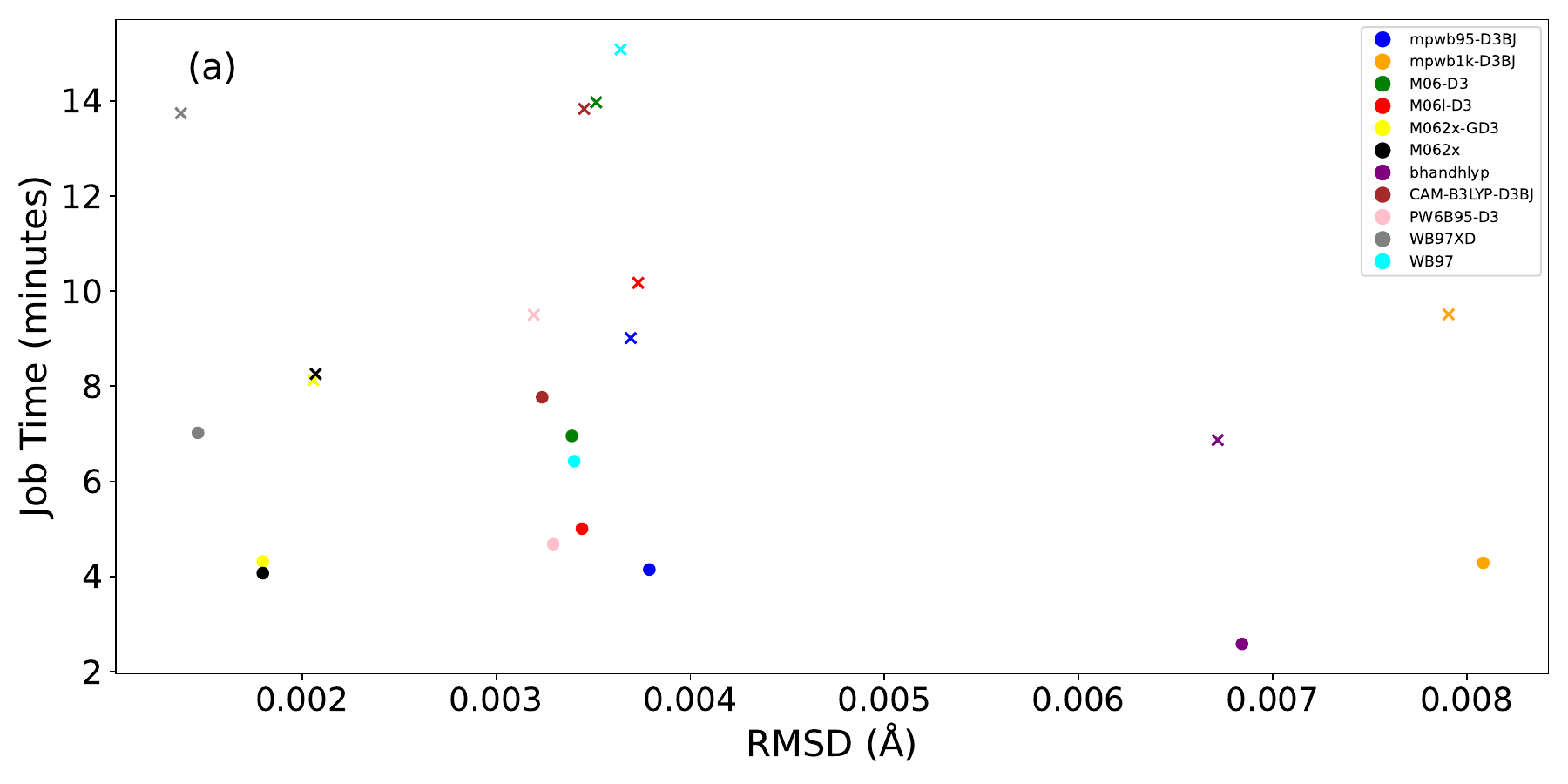}
    \vspace{0.3cm}
    \includegraphics[height=0.35\textheight]{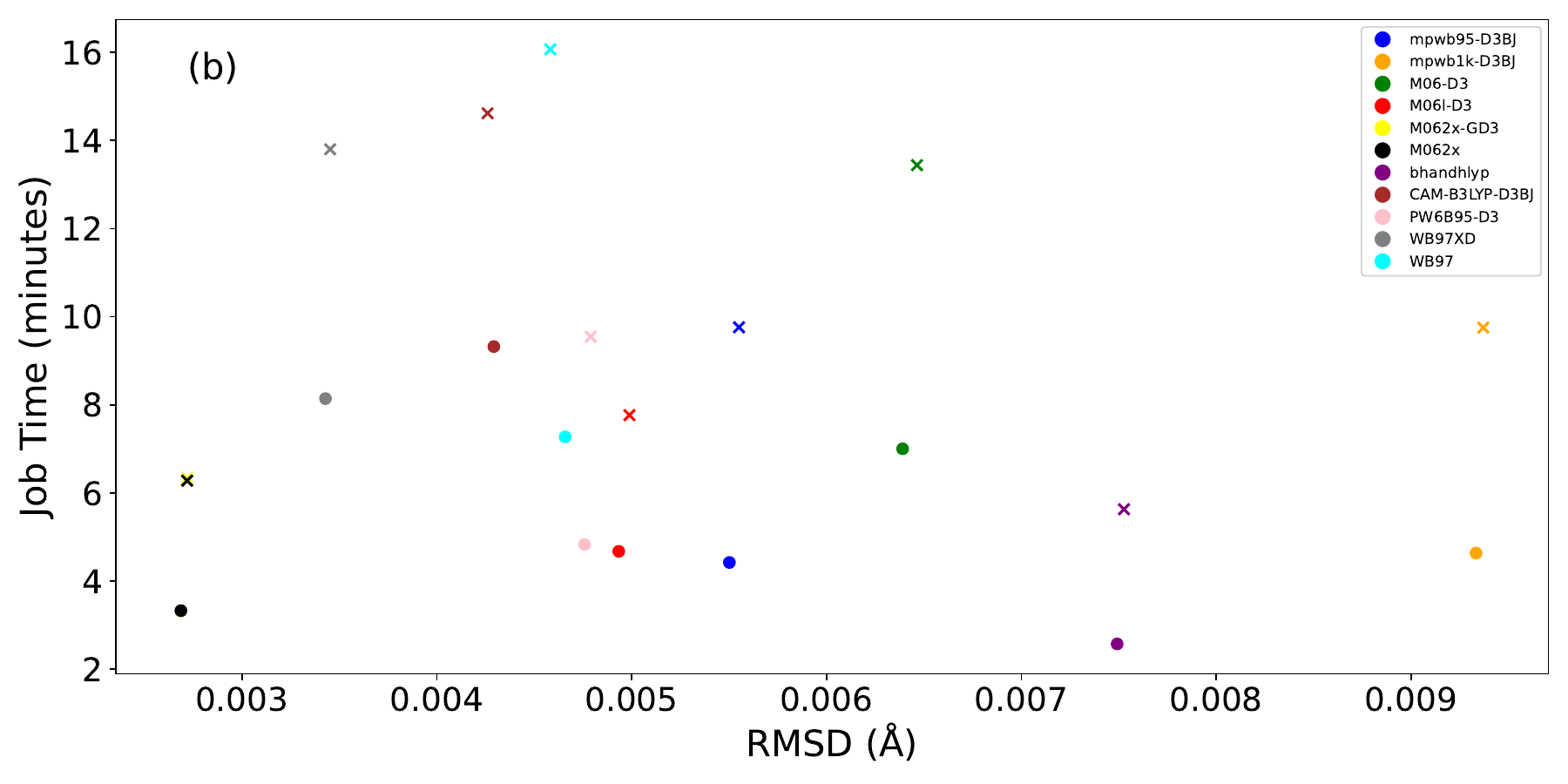}
    \caption{RMSD in bond length (\text{\AA}) for different xc-functionals used in the geometry optimization for step 1, (a), and step 2, (b), in the gas phase at def2-TZVP(O) and def2-TZVPD(X). The reference geometry is DF-CCSD(T)-F12/cc-pVDZ-F12.}
    \label{fig:benchmarking}
\end{figure}

\begin{figure}[ht]
    \centering
    \includegraphics[width=0.35\textheight]{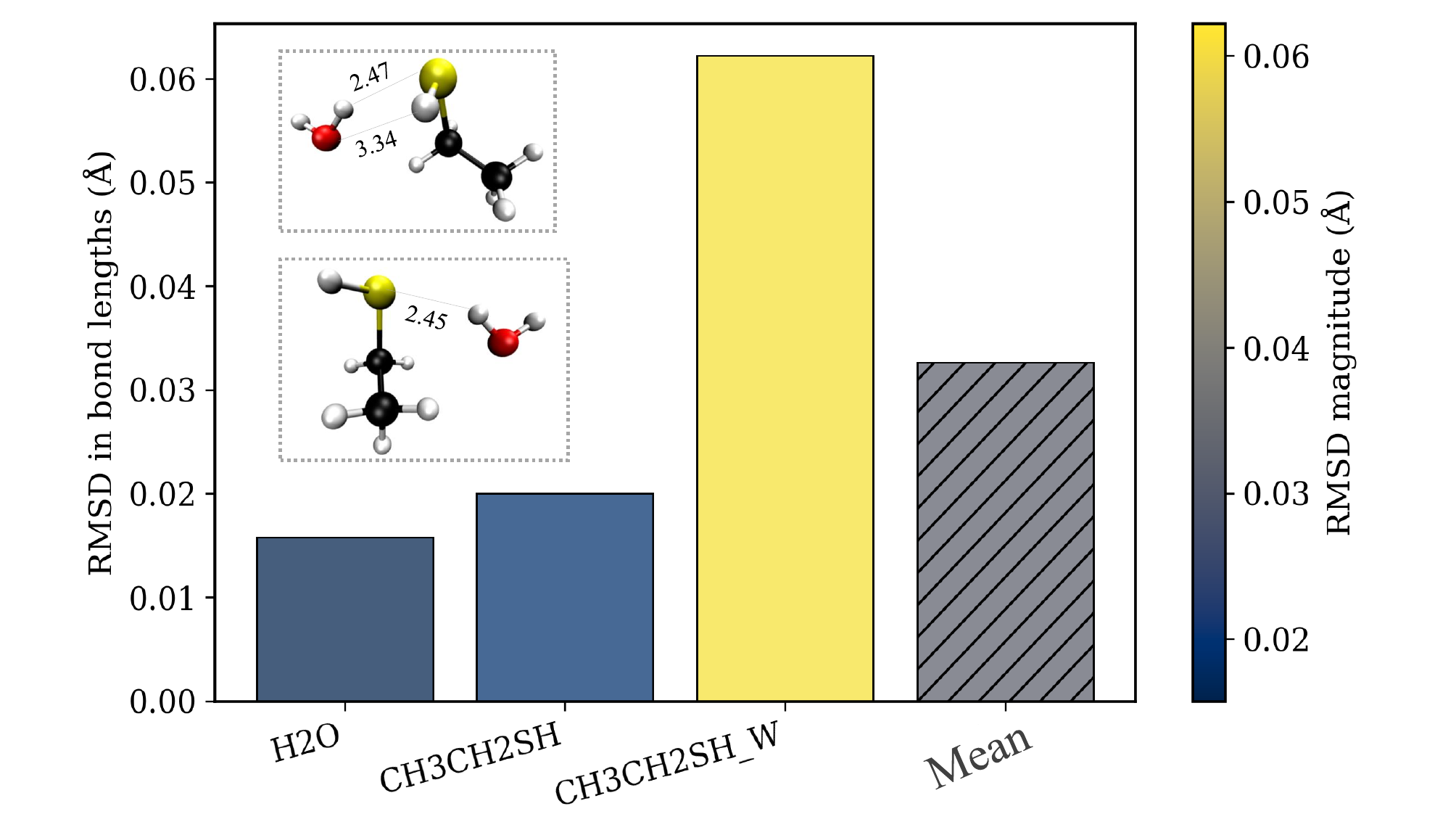}
    \caption{\textcolor{black}{RMSD in bond lengths between between HF-3c/MINIX and reference M06-2X/def2-TZVP optimized geometries for H$_2$O, CH$_3$CH$_2$SH, and CH$_3$CH$_2$SH adsorbed on a single water molecule (W1) in two different orientations (combined as CH$_3$CH$_2$SH\_W), The inset depicts the two orientations, and the final bar represents the mean RMSD across all systems.}}
    \label{fig:benchmarkinghf3c}
\end{figure}

\begin{figure}[ht]
    \centering
    \includegraphics[width=0.80\textwidth]{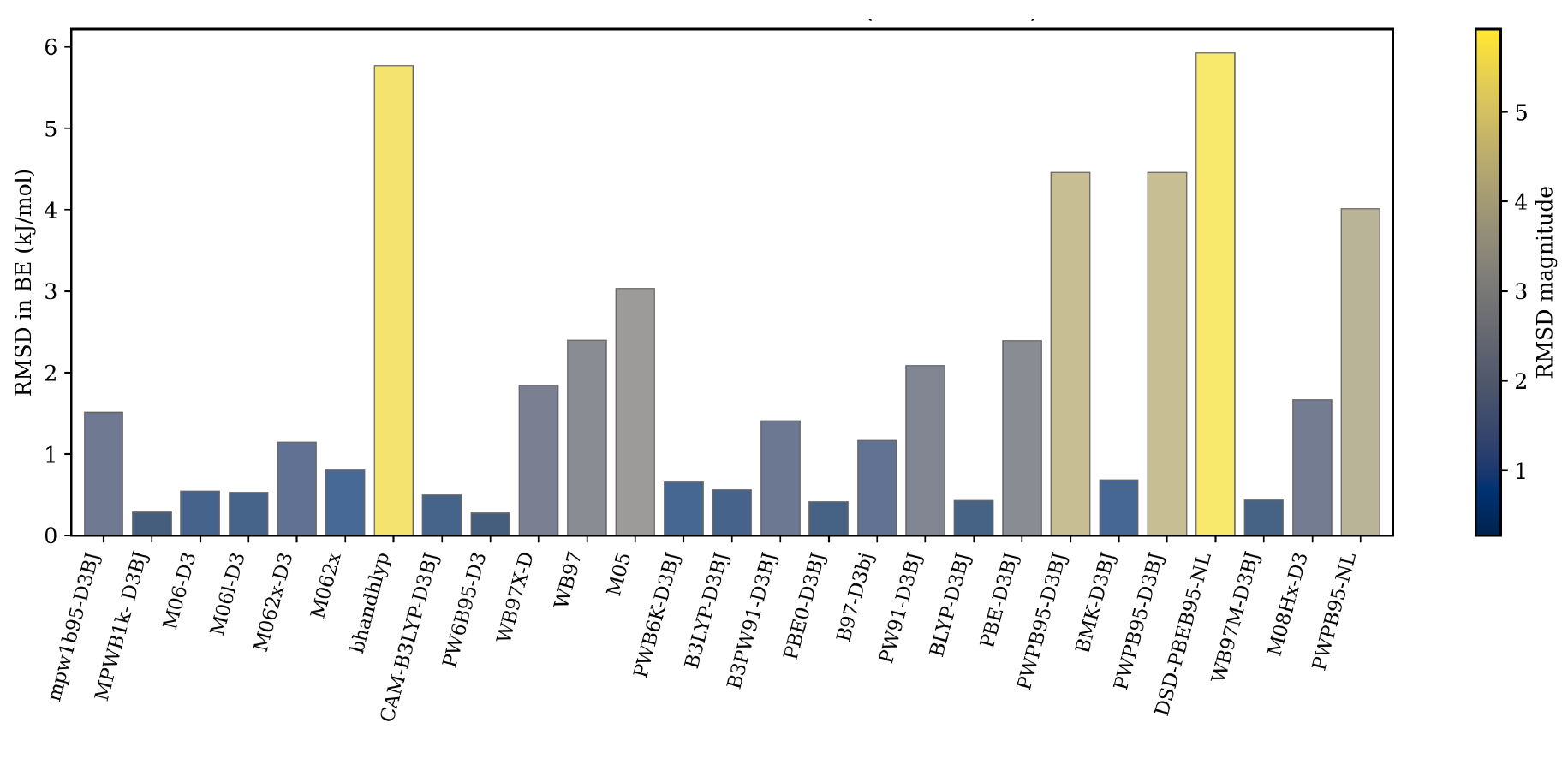}
    \caption{\textcolor{black}{Average RMSD in binding energy (BE) for the CH$_3$CH$_2$SH-W1 complex (for two orientations as in Fig.~\ref{fig:benchmarkinghf3c}) with the def2-TZVPD basis set. The RMSD is computed with respect to CCSD(T)/CBS//M06-2X/def2-TZVP reference binding energies}.}
    \label{fig:benchmarkingW1}
\end{figure}

\begin{figure}[ht]
    \centering
    \includegraphics[width=0.90\textwidth]{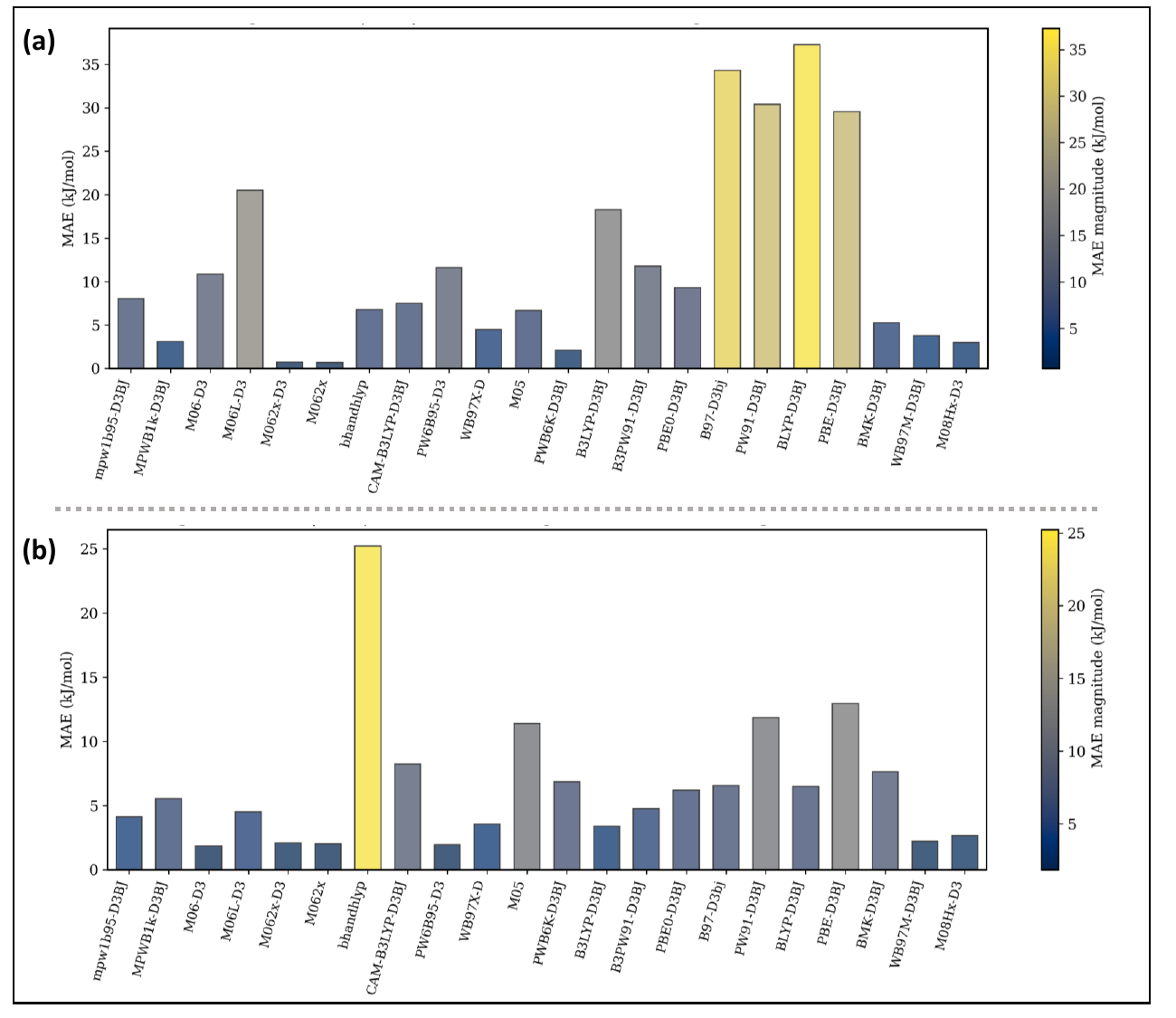}
    \caption{\textcolor{black}{Mean absolute error (MAE) for different xc-functionals in (a) activation energy $\Delta E^{\ddagger}$ and (b) reaction energy $\Delta E$, for the Step 1 and Steb 2(a)-2(b) in the gas phase. The MAE is computed with respect to CCSD(T)/CBS//M06-2X/def2-TZVP reference energies}.}
    \label{fig:bnchmark_rxnEnergy}
\end{figure}
To accurately explore the reaction potential energy surface both in the gas phase and on ice grain surfaces for the studied reaction scheme, we first conducted a benchmark study to identify ~\textcolor{black}{the suitable electronic-structure method:}\\
(i) To select appropriate DFT method for geometry optimization, the DF-CCSD(T)-F12/cc-pVDZ-F12 level of theory is used as a reference. The geometries of the reactants and products involved in reactions (1) and (2) are optimized using eight common exchange-correlation functionals. The average root mean square deviation (RMSD) of these geometries, computed with two different basis sets, def2-TZVP and def2-TZVPD, is presented in Fig.~\ref{fig:benchmarking}, along with the associated computational time. All the tested functionals yielded satisfactory results for the current reaction system, with average RMSD values below 0.001 Å. Among them, the WB97X-D functional demonstrated superior performance in terms of geometry accuracy \textcolor{black}{for step 1 and M06-2X performed best for step 2}, while BHandHLYP is the most time-efficient approach. Notably, the inclusion of dispersion corrections in M06-2X-D3 did not result in significant geometric improvements compared to WB97X-D. However, \textcolor{black}{M06-2X functional offered an optimal balance between accuracy and computational efficiency, delivering reliable geometries while minimizing computational time for both the steps}. Furthermore, the addition of diffuse functions, def2-TZVPD provided no substantial enhancement in geometry, but did incur a significant increase in computational cost. \textcolor{black}{Therefore, preliminary PES search for both gas-phase and on ice grain-surface is performed at cost-effective BHandHLYP/def2-TZVP followed by geometry refinement at M06-2X/def2-TZVP. }\\

\textcolor{black}{(ii) In addition, to validate the structural quality of HF-3c/MINIX, used to refine the geometries of all the binding sites on W22, we compared the HF-3c/MINIX bond lengths against those obtained from M06-2X/def2-TZVP optimizations for representative systems: H$_2$O, CH$_3$CH$_2$SH, and CH$_3$CH$_2$SH adsorbed on a single water molecule (W1) in two different orientations. The RMSD in bond lengths is found to be ($\leq 0.02$--$0.06$~\AA; Fig. \ref{fig:benchmarkinghf3c}(b)), demonstrating that HF-3c/MINIX geometries are sufficiently accurate for describing hydrogen-bonded configurations.\\}

(iii) Further, we benchmarked 27 hybrid meta-GGA and GGA functionals with the def2-TZVPD basis set by comparing the binding energies of the CH$_3$CH$_2$SH–W1 complex (in two different orientations) against CCSD(T)/CBS reference values (Fig.~\ref{fig:benchmarkingW1}). PW6B95-D3BJ and MPWB1K-D3(BJ) functional, showed good performance for this system. Beyond this, based on the results from our ongoing benchmarking analysis for broader sets of interstellar molecules, MPWB1K-D3(BJ) appears to be an overall reliable functional.\\

\textcolor{black}{(iv) Finally, the mean absolute error (MAE) for the different xc-functionals is computed for activation energy $\Delta E^{\ddagger}$ and (b) reaction energy $\Delta E$, in the gas phase with respect to CCSD(T)/CBS//M06-2X/def2-TZVP reference energies \ref{fig:bnchmark_rxnEnergy}, to choose appropriate DFT functional for the single-point energy refinement of the ice-grains mechanisms.}\textcolor{black}{The functional MPWB1K-D3(BJ) performed well for reaction energetics with mean absolute error (MAE) $\leq$ 3.3 kJ/mol for $\Delta E^{\ddagger}$ and MAE $\leq$ 5.5 kJ/mol for $\Delta E$, Fig.\ref{fig:bnchmark_rxnEnergy}. As it also showed superior performance for binding energy, Fig.\ref{fig:benchmarkingW1} so MPWB1K-D3(BJ)/def2-TZVPD//M06-2X/def2-TZVP was used for the final reaction energy calculations at ice-surfaces.} \\

For geometry optimizations at the CCSD(T)-F12 level, the MOLPRO program is employed \cite{werner2012molpro}, while all other DFT optimizations are performed using Gaussian 16. The geometry optimizations at HF-3c/MINIX and all the energy calculations are conducted using the Psi4 program \cite{turney2012psi4}.\\
\textcolor{black}{Notably, the choice of different functional tested in Fig. \ref{fig:benchmarking}, \ref{fig:benchmarkingW1} and \ref{fig:bnchmark_rxnEnergy}, are selectively chosen from an extensive benchmarking study performed in the previous work of the group \cite{bovolenta2024depth}.}

\section{Potenitial Energy Surface Scan}\label{appendix:ScanSection}

\begin{figure}[ht]
    \centering
    \includegraphics[width=0.70\textwidth]{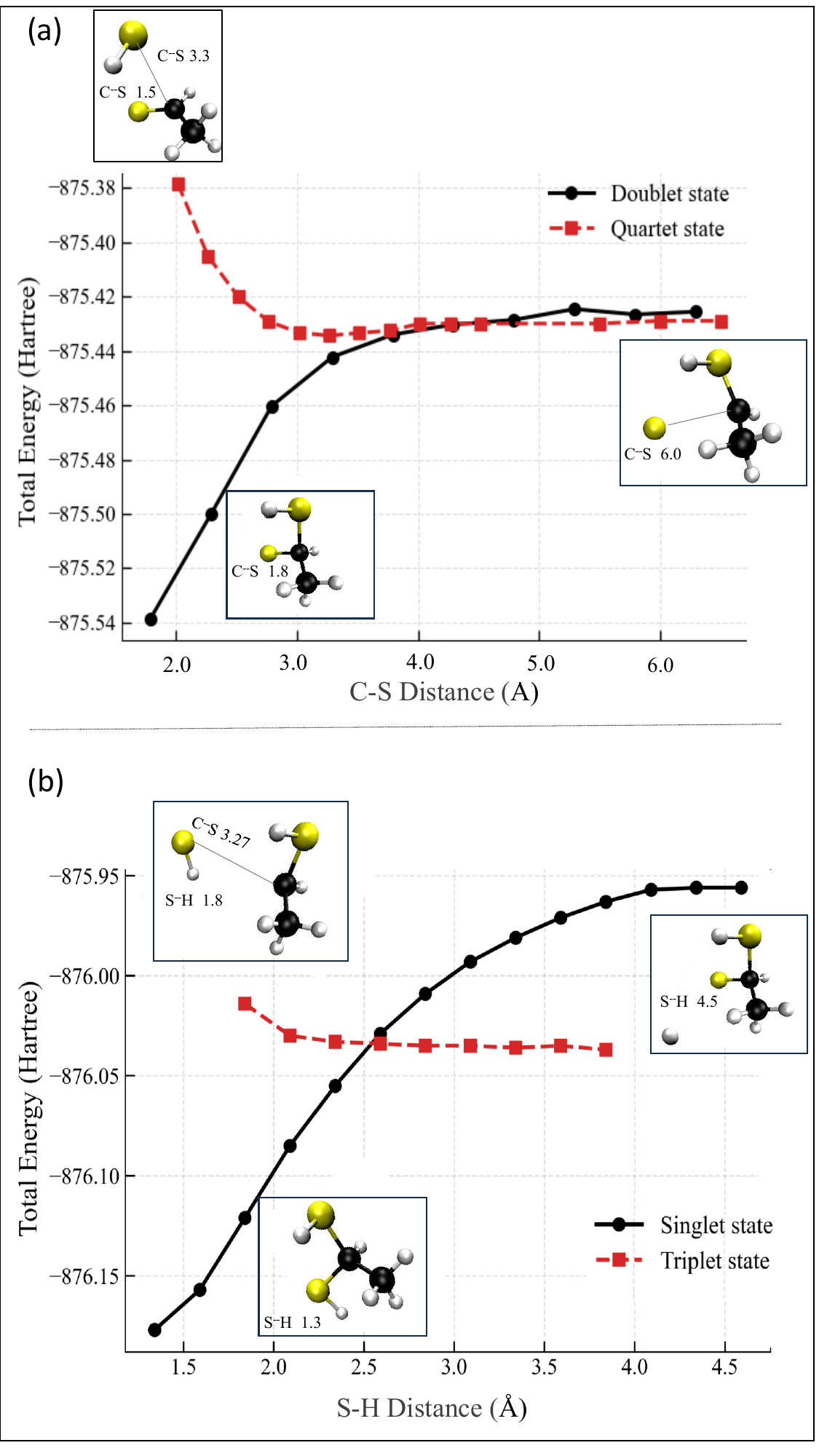}
    \caption{\textcolor{black}{Relaxed potential energy scans at M06-2X/def2-TZVP comparing different spin states as indicated along the (a) C–S distances for Step 2(a) CH3CHSH + S; and (b) S–H distances for Step 2(c) CH3CH(S)SH + H.}}
    \label{fig:scan}
\end{figure}

\textcolor{black}{The relaxed potential energy scans at the low-spin and high-spin electronic states employing M06-2X/def2-TZVP is depicted in Fig.~\ref{fig:scan}. The scans were initiated from physically separated radical fragments and continued toward shorter separations along the reaction coordinate. At the high spin surface, the interaction is clearly repulsive at short distances and relaxes back toward a weakly bound configuration at longer distances, consistent with a non-reactive high-spin surface. In contrast, on the low spin surface, the energy decreases monotonically as the distance shortens, with no intervening maximum or saddle point between the separated radicals and in the bonded product region. This indicates that the association proceeds without a detectable barrier along the reaction coordinate. Further, for high spin surfaces the system starts fragmenting below 2.25~\AA{}, both for quartet, Fig.~\ref{fig:scan}(a) and triplet states, Fig.~\ref{fig:scan}(b). Interestingly, the location of the intersystem crossing (ISC) beyond the bonding region suggests that product formation may occur through population transfer from the high-spin to the low-spin potential energy surface, followed by a rapid and strongly exothermic relaxation into the product well. A quantitative description of this nonadiabatic mechanism is, however, beyond the scope of this study.}

\end{appendix}
\twocolumn

\end{document}